\documentclass[epjST]{svjour}
\usepackage{graphics}
\usepackage{color}
\usepackage{graphicx}
\usepackage{amsmath,amssymb,amsfonts}

\usepackage{caption}
\newcommand{\out}[1]{}

\newcommand{\ii}{\mathrm{i}}
\newcommand{\ee}{\mathrm{e}}

\newcommand{\rr}{\mathbf{r}}
\newcommand{\sg}{\mathbf{g}}

\newcommand{\UG}{G}
\newcommand{\OG}{G}

\definecolor{orange}{rgb}{1,0.5,0}

\begin{document}
\title{Merging {\it \textbf{GW}} with DMFT and non-local correlations beyond}

   \author{J. M. Tomczak\inst{1} \and  P. Liu\inst{2,3} \and  A. Toschi\inst{1} \and  G. Kresse\inst{2} \and K. Held\inst{1}\fnmsep\thanks{\email{held@ifp.tuwien.ac.at}}
}
  \institute{
  Institute of Solid State Physics, TU Wien,  1040 Wien, Austria  \and
    University of Vienna, Faculty of Physics and Center for
Computational Materials Science, Sensengasse 8/12, A-1090 Vienna, Austria \and
Shenyang National Laboratory for Materials Science, Institute of Metal Research,
University of Chinese Academy of Sciences, Shenyang 110016, China
}
  \abstract{We review recent developments in electronic structure calculations that go beyond state-of-the-art methods such as density functional theory (DFT) and dynamical mean field theory (DMFT). Specifically, we discuss the following methods: {\it GW} as implemented  in the Vienna {\it ab initio} simulation package (VASP) with the self energy on the imaginary frequency axis, {\it GW}+DMFT, and {\em ab initio} dynamical vertex approximation (D$\Gamma$A). The latter includes the physics of {\it GW}, DMFT and non-local correlations beyond, and allows for calculating (quantum) critical exponents. We present results obtained by the three methods with a focus on the benchmark material SrVO$_3$.
    %
    %
  }

\maketitle


\section{Introduction}
\label{Sec:Int}
The calculation of materials with predictive power is arguably the biggest challenge of condensed matter theory. In the 20th century we have seen the breakthrough of density functional  theory (DFT) \cite{Kohn1965,Hohenberg1964} (for reviews see Refs.~\cite{Jones1989a,Martin04}) which allows for the reliable calculation of many materials and their properties. This is quite surprising considering the fact that the approximations employed to the exchange and correlation potential, such as the local density approximation (LDA) or the generalized gradient approximation (GGA), are rather crude.
Despite the success of DFT for many materials, there are entire classes of systems for which it does not work properly. This happens, e.g.,  for materials,
in which exchange or correlation effects are large. Hence the silver bullet of method development is to find better potentials or to improve upon exchange and correlations by many-body methods
\cite{Martin16}.

\begin{figure}
  \begin{center}   \includegraphics[width=0.7\columnwidth,clip]{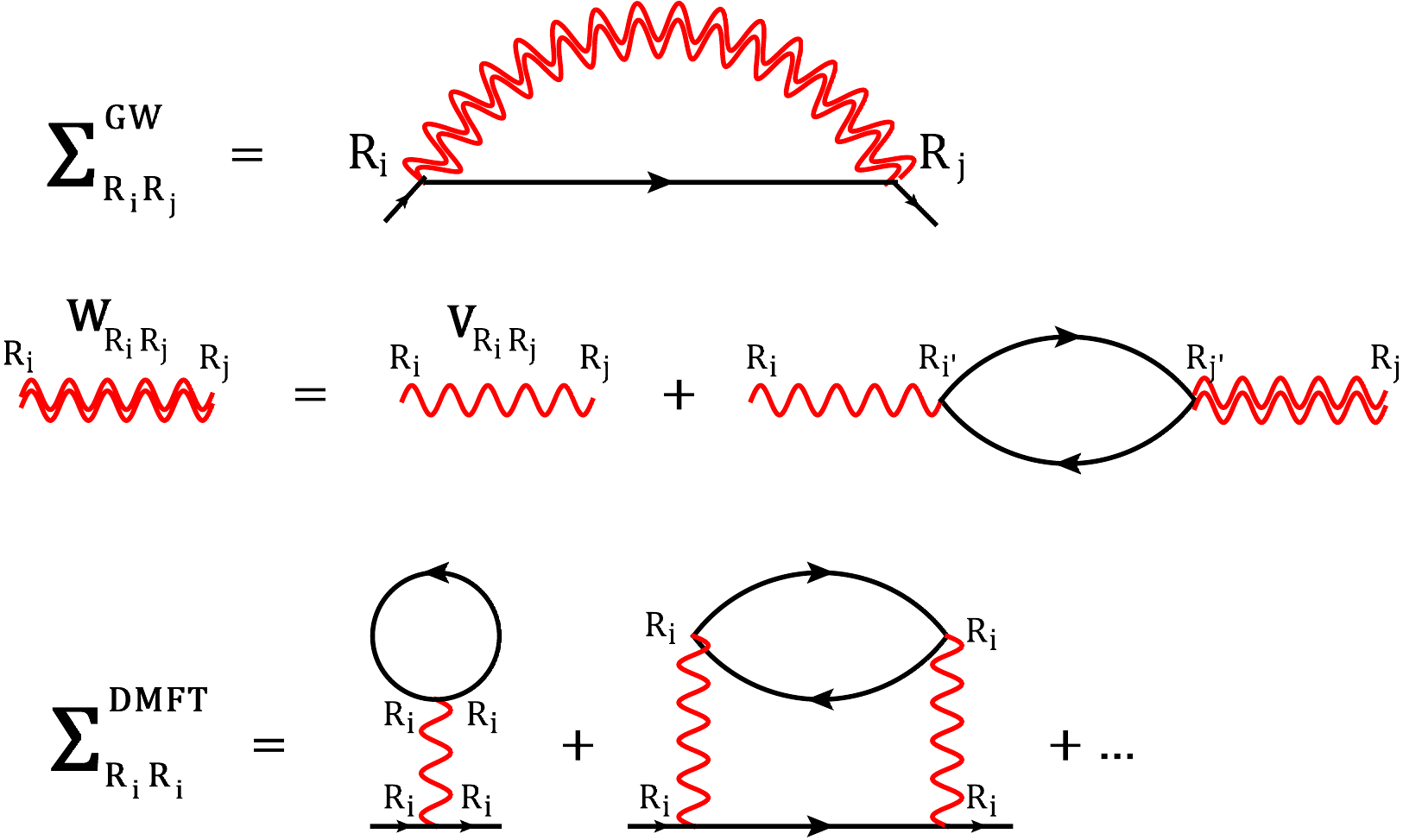}
\vspace{.5cm}

\includegraphics[width=0.85\columnwidth,clip]{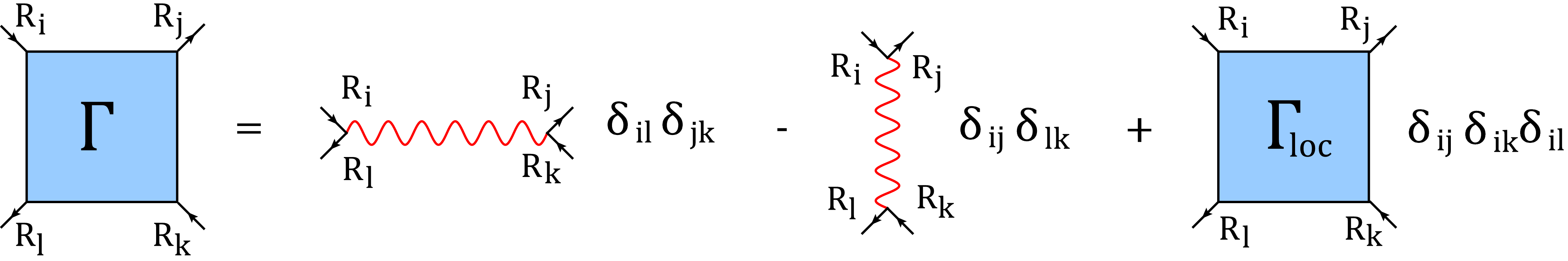} 
    \end{center}
    \caption{\label{Fig1}
		{\bf First line:} In $GW$ the self energy $\Sigma$ is given by the interacting Green function $G$ (black straight line) times the screened interaction $W$ (red wiggled  line) from coordinate/site $R_i$ to $R_j$.
		 {\bf Second line:} The screened interaction $W$ in turn is given by the bare interaction (here denoted as $V$) and the screening in the random phase approximation (RPA). This RPA screening is generated by the last term which yields a ladder in terms of $V$ and bubbles consisting of two Green functions.
 {\bf Third line:} In DMFT the self energy is given by the local contribution of {\em all} Feynman diagrams with the local interaction $U$ always on the same site $R_i$. {\bf Fourth line:} In AbinitioD$\Gamma$A, we take as the irreducible vertex $\Gamma$ the bare non-local Coulomb interaction $V^q$ and the local vertex $\Gamma_{\rm loc}$ which depends on orbitals ($l$, $m$ ...)  and  frequencies ($\nu$, $\nu'$, $\omega$) but not momenta ($k$,$k'$,$q$);  $\Gamma_{\rm loc}$ also includes the local Coulomb interaction $U$
(adapted from Ref.~\cite{Galler2016}).  }
  \end{figure}

Materials in which the exchange part is particularly important are, e.g., semiconductors. Here, DFT within LDA or GGA predicts consistently too small band gaps. This can be overcome by hybrid functionals \cite{Seidel1996,Adomo1999,Heyd2003,Fuchs2007} that mix part of the exact exchange to the exchange correlation functional.
The amount of exact exchange that is required for an accurate modeling is, 
however, non-universal, i.e., material-dependent. For instance, in metals
the  long-range exchange is screened by long-range charge fluctuations \cite{Freysoldt_defects_RevModPhys_2014}.
An accurate  many-body framework to capture the system-dependent screening is Hedin's  {\it GW} approach \cite{Hedin1965}
which calculates the screened-exchange self energy from the Green function $G$ times the screened exchange $W$, see Fig. \ref{Fig1} for the corresponding Feynman diagram.
Most $GW$ results have been obtained using a DFT-derived Green function $G_0$ and an interaction $W_0$ that has been screened by the Lindhard function computed with $G_0$.
Only recently self-consistent $GW$ calculations that use an approximate hermitianized form of the self energy, as proposed by  van Schilfgaarde and Kotani \cite{Faleev2004,Chantis2006}, became available. In {\bf Section \ref{Sec:GW}} we discuss the $GW$ method and the calculation of the full frequency-dependence of the self energy, which is needed for spectral functions and for a self-consistency beyond the van Schilfgaarde--Kotani approximation. We detail in particular the advantages of our new imaginary-frequency implementation of $GW$ within the Vienna {\it ab initio} simulation package (VASP) \cite{Shishkin2006,Liu2016}.
Both methods, hybrid functionals and {\it GW},  can  lead to semiconductor band gaps in far better agreement with experiment \cite{Godby1988,Seidel1996,Fuchs2007},
with the $GW$ self energy  acting as a ``scissors operator'' \cite{Godby1988}.
Beyond that, $GW$ also describes quasiparticle renormalizations, finite life times, and improves on the total energies of, e.g., defects \cite{Aryasetiawan1998,Freysoldt_defects_RevModPhys_2014}.
While hybrid functionals are one-electron-like by construction, also the {\it GW}---at least in all common implementations (see however the recent Refs~\cite{PhysRevB.94.155101,PhysRevB.95.035139})---is based on a
Green function $G_0$ that is always related to a single Slater determinant.
Excluding any multi-reference character in $G_0$, the $G_0W_0$ approach is thus not capable to treat systems in which fluctuations are strong.

Materials in which  the correlation part is particularly important are, among others, transition metal oxides and heavy fermion compounds with partially filled $d$ and $f$ shells, respectively \cite{Imada1998}. For treating electronic correlations in such materials, dynamical mean field theory (DMFT) \cite{Metzner1989,Georges1992a} (for a review see Ref.~\cite{Georges1996})
and its merger with DFT \cite{Lichtenstein1998,Anisimov1997} (for reviews see Refs.~\cite{Kotliar06,Held2007}) has been a big leap forward.
DMFT takes into account a major part of the electronic correlations: the ``local'' ones that are confined to a single atomic site. Fig. \ref{Fig1} (bottom) shows the corresponding Feynman diagrams.
This way, among others, quasiparticle renormalizations including kinks \cite{Nekrasov05a,Byczuk2007,Toschi2009,Held13}, Hubbard side bands, metal-insulator transitions, and magnetism can be described much more accurately than
with one-particle methods, and finite temperature properties become accessible as well. Early successes of DFT+DMFT include the calculation of the Mott-Hubbard transition in V$_2$O$_3$ \cite{Keller2004,Held2001a,Poteryaev2007,Hansmann2013b}, magnetism in Fe and Ni \cite {Lichtenstein1998}, and the $\alpha$-$\gamma$ transition in Ce \cite{Held2001,McMahan2003}. More recently, it has also been applied to oxide heterostructures \cite{Okamoto2005,Hansmann2009,PhysRevLett.114.246401}, surfaces \cite{PhysRevLett.115.236802}, nanoclusters \cite{Valli2015a} and oxygen vacancies \cite{PhysRevB.94.241110}.

The major remaining shortcomings of DFT+DMFT are (i) the sand in the clockwork when interfacing a density functional theory with a Feynman diagrammatic approach and
(ii) that only local correlations are taken into account in DMFT.
Regarding (i), let us in particular mention the double counting: It is unclear which part of the DMFT correlations are taken into account already on the DFT side, and so different double counting schemes have been proposed. Most commonly used is the fully localized limit \cite{Anisimov1991}. The double counting issue is  particularly pronounced in so-called ``$d+p$'' DFT+DMFT calculations that in, say, oxides, include both, the transition metal $d$- as well as the oxygen $p$-orbitals and can lead to largely different results \cite{Hansmann2014,Dang2014,Haule2014}.

This conceptual problem can be overcome by substituting DFT by $GW$ in the so-called {\it GW}+DMFT approach \cite{Sun02,Biermann2003}, which merges two many-body Feynman diagrammatic approaches so that one can  precisely identify which diagrammatic contribution is counted twice. {\it GW}+DMFT also provides for a better treatment of the exchange contribution. This is not only of advantage for correlated semiconductors such as Ga$_{1-x}$Mn$_x$As, and ligand-states in, e.g., transition metal oxides \cite{Tomczak12}, but also for a quantitative description of effective masses of correlated electrons \cite{Tomczak14}.
We discuss the {\it GW}+DMFT approach and present results in {\bf Section \ref{Sec:GWDMFT}}; for a more detailed introduction we refer the reader to Refs.~\cite{Held2011,gwdmft_proc1,0953-8984-26-17-173202,0953-8984-28-38-383001}.
One should note, however, that the treatment of non-local correlations is very limited in {\it GW}+DMFT as only charge fluctuations and only the particle-hole channel are included in {\it GW}. Moreover they are treated  only in weak coupling perturbation theory, i.e., by building the particle-hole ladder only in terms of the bare Coulomb interaction $V$, see Fig.\ \ref{Fig1} (middle).

There are essentially two routes that deal with non-local correlations while keeping the local DMFT correlations at the same time: cluster \cite{Hettler1998,Lichtenstein2000,Kotliar2001} and diagrammatic extensions \cite{Toschi07,Kusunose06,Rubtsov2008,Rohringer2013,Taranto2014,Ayral2015,Li2015} of DMFT.
The former have been successfully applied to the two dimensional Hubbard model and helped establishing the presence of superconductivity in this model. However, due to numerical restrictions, realistic multi-orbital calculations are only possibly for a handful of sites, restricting the cluster extensions essentially to nearest neighbor correlations (for a review see Ref.~\cite{Maier05}).

Diagrammatic extensions of DMFT on the other hand can treat short- and long-range correlations on an equal footing, which allowed, among others, the calculation
of critical exponents \cite{Rohringer2011,Hirschmeier2015,Antipov2014,Schaefer2016} and revealed the absence of a metal-insulator transition in the two-dimensional Hubbard model on a square lattice \cite{Schaefer2015-2}.
In {\bf Section \ref{Sec:DGA}} we discuss the first of these diagrammatic extensions, the dynamical vertex approximation (D$\Gamma$A) \cite{Toschi07,Katanin2009}
and its extension to {{\em ab initio} calculations.
For the latter, AbinitioD$\Gamma$A \cite{Toschi2011,Galler2016}, we take as  the  vertex (irreducible in the particle-hole channel)  the bare non-local Coulomb interaction as well as the local Coulomb interaction and all local vertex diagrams, see Fig.\ \ref{Fig:DGA}. From this unifying framework, we naturally generate all (local and non-local) {\it GW} diagrams, all local DMFT diagrams, as well as non-local diagrams beyond. The latter include, e.g.,  spin fluctuations which are
important in the vicinity of phase transitions, for magnons and pseudogap physics. For a pedagogical introduction see Ref.~\cite{Held2014}, and  Ref.~\cite{Georgthesis2013} for an elaborate presentation.


\section{Hedin's {\it GW} method: The new VASP implementation}
\label{Sec:GW}

\subsection{Method}

Hedin's method is in principle an exact approach to describe many-body interactions \cite{Hedin1965,strinati1980,hybertsen1986,Aryasetiawan1998,bechstedt2009}. However, in practice for computational
reasons, virtually all implementations of this method are limited to the so-called $GW$ approximation. This greatly simplifies the calculations,
but also makes important approximations\footnote{Only the particle-hole channel is considered and the vertex is approximated by the bare Coulomb interaction $V^q$, see Section \ref{Sec:DGA}.}; the considered Feynman diagrams are shown in the top panel of Fig. \ref{Fig1}.

The new aspect of the present VASP implementation \cite{Liu2016} is that it is tuned for massively parallel computers and
that it works in imaginary time and frequency as opposed
to the earlier VASP implementation that worked along the real frequency axis \cite{Shishkin2006,ShishkinPRL2007}
and necessitated very fine frequency grids.
In the following, we will give a brief outline of the computational steps of the present code, highlighting why it
is particularly convenient for a combination with DMFT.  We follow previous publications but emphasize simplicity and conciseness
by dropping for instance the Brillouin zone index as well as the PAW formalism~\cite{Liu2016}.
The first step in a $GW$ calculation is to determine the DFT one-electron orbitals $\psi_i$
and one-electron energies $\epsilon_i$. From the DFT orbitals the one-electron Green function follows:
\begin{eqnarray}
\label{eq:Green_occ}
\OG(\rr,\rr', \ii\tau) =& \phantom{-} \sum\limits^{\text{\phantom{u}occ\phantom{n}}}\limits_i \psi_i(\rr)
\psi^*_i(\rr') \ee^{-(\epsilon_i-\mu )\tau} \quad (\tau<0) ,  \\
\label{eq:Green_unocc}
\UG(\rr,\rr', \ii\tau)=&-\sum\limits^{\text{unocc}}\limits_a \psi_a(\rr)
\psi^*_a(\rr') \ee^{-(\epsilon_a-\mu) \tau} \quad (\tau>0).
\end{eqnarray}
Generalization to finite temperature is straightforward and involves restriction of the time
to $-\beta \le \tau \le \beta$, where $\beta$ is the inverse temperature, and introduction of  Fermi occupancy
factors   $n_i= 1/(\exp((\epsilon_i-\mu) \beta)+1)$ and $(1-n_i)$ in the first and second equation, respectively.
It is then easy to show that the function observes the anti-periodicity for Fermionic Green functions
$ \UG(\rr,\rr', \ii\tau) = - \OG(\rr,\rr', \ii (\tau-\beta))$.

As typically done in plane wave codes, all functions are expanded in a plane wave basis and fast Fourier transformed (FFT)
to real space only when this is required:
\begin{eqnarray}
\label{eq:FFT_G1}
G(\rr, \rr', \ii\tau) & = &
\sum\limits_{\sg, \sg'} \mathrm{e}^{\ii  \sg \rr} G (\sg,\sg', \ii\tau)  \mathrm{e}^{-\ii\sg'\rr'}\\
G(\sg, \sg', \ii\tau) & = & \frac{1}{N_{\bf r}^2}
\sum\limits_{\rr, \rr'} \mathrm{e}^{-\ii  \sg \rr} G (\rr,\rr', \ii\tau)  \mathrm{e}^{\ii\sg'\rr'}.
\end{eqnarray}
Here $N_{\bf r}$ is the total number of real-space grid points. Since the plane-wave basis can be chosen to be significantly
smaller than the number of real space grid points \cite{Payne_RevModPhys_1992},
 the plane wave expansion  typically reduces the storage demand by a factor 6-8 for orbitals (one position index), and a
factor $6^2-8^2$ for Green functions (two position indices). The first crucial approximation
of the $GW$ method is that the irreducible polarizability is approximated by the  independent particle polarizability
(RPA). Assuming a factor 2 for spin-degenerate systems, we get
\begin{equation}\label{eq:chi_GG}
 P(\rr,\rr',\ii\tau)=2 G(\rr,\rr',\ii\tau)G(\rr',\rr,-\ii\tau)= 2 G(\rr,\rr',\ii\tau)G^*(\rr,\rr',-\ii\tau),
\end{equation}
that is, vertex corrections of the form $P=2 GG\Gamma$ are neglected. This approximation neglects important many body effects, for instance
excitonic effects \cite{ShishkinPRL2007,Freysoldt_defects_RevModPhys_2014}
that are captured by particle-hole ladder diagrams. It has been shown that these terms become important
when selfconsistent calculations are performed \cite{ShishkinPRL2007}.
From the irreducible polarizability the  screened interaction (see Fig. \ref{Fig1} middle)
can be determined by:
\begin{equation}
\label{eq:W}
 W({\bf r}, {\bf r}', \ii\omega)= V({\bf r}, {\bf r}') +V({\bf r}, {\bf s}) P({\bf s}, {\bf s}', \ii \omega) W({\bf s}', {\bf r}', \ii\omega) \; \Leftrightarrow \; W^{-1} = V^{-1} - P.
\end{equation}
Here, $V$ is the Coulomb kernel, and integration over repeated spatial coordinates ($\bf s$ and ${\bf s}'$) is assumed.
For reasons of computational efficiency, the calculation is more conveniently done in reciprocal space, where
the Coulomb kernel is diagonal \cite{Liu2016}.

The  Dyson-like equation for the screened interaction needs to be solved in
frequency space $\ii\omega$. This obviously requires one to perform a Fourier transformation of the
independent particle polarizability from imaginary time [compare Eq. (\ref{eq:chi_GG})] to imaginary frequency. In previous (imaginary time) $GW$ codes \cite{Godby1995,Godby2000}
this was a fairly cumbersome
operation involving fitting, a fast Fourier transformation, and some analytic continuation at very large frequencies and times.
Using a mathematical rigorous treatment, Kaltak {\em et al.} determined imaginary time and frequency
grids \cite{KaltakJCTC2014} that have a number of favorable properties.  (i) The grids are non-uniformly spaced. This  allows  to simultaneously and accurately describe intra-band
transitions at very small energies (meV),
as well as high energy excitations into continuum like  states (up to several 100 eV). (ii) The time and frequency grids are individually  optimized to
allow accurate calculations of the correlation energy in  second order. Convergence of the correlation energy is exponential in the number of time or frequency points, with 20 points yielding $\mu$eV convergence even for metals. (iii) The grids are dual to each other:
if a Bosonic  function is known at a grid of $N_{\omega}$  frequency points $\omega_k, k=1,...,N_\omega$, the numerical error in the Bosonic
function is minimal at a set of corresponding $N_\tau=N_\omega$ imaginary time points $\tau_j, j=1,...,N_\tau$. (iv) Related to point (iii), a numerical
discrete Fourier transformation exists to transform any function from imaginary time to imaginary frequency (and vice versa):
\begin{equation}
\label{eq:cos_sin_sigma1}
f(\ii \omega_k)  =
\sum\limits_{j=1}^{N_\omega}\gamma_{kj} \mathrm{cos}(\omega_k \tau_j) (f(\ii\tau_j)+f(-\ii\tau_j)) +
 \ii \, \lambda_{kj}\mathrm{sin}(\omega_k \tau_j)  (f(\ii\tau_j)-f(-\ii\tau_j)).
\end{equation}
This is a numerical approximation to the Fourier transformation from time to frequency :
\begin{eqnarray}
 f(\ii \omega) &= &\int_{-\infty}^{\infty}  f(\ii \tau) e^{\ii \omega \tau} d \tau \nonumber \\
  &=& \int_{0}^{\infty} \Big[ \cos( \omega\tau )(f(\ii \tau)+f(-\ii \tau)) +\ii \sin(  \omega\tau) (f( \ii \tau)-f(-\ii \tau)) \Big]d \tau.
\end{eqnarray}
The corresponding matrix of coefficients, e.g., $\gamma_{kj} \mathrm{cos}(\omega_k \tau_j)$, are precalculated and
stored.

The evaluation of the self energy is most conveniently done in imaginary time
\begin{equation}
\label{eu:selfenergy}
  \Sigma^c({\bf r}, {\bf r}',  \ii \tau) = -G({\bf r}, {\bf r}', \ii\tau)  W^c({\bf r}, {\bf r}', \ii\tau), \qquad W^c(\ii\omega)= W(\ii \omega)-V \;
\end{equation}
with the bare Coulomb kernel $V$ subtracted before the Fourier transformation
of $W$ and the contribution $G V$ calculated analytically.
As for the polarizability, also Eq. (\ref{eu:selfenergy}) neglects vertex corrections ($\Sigma=-GW\Gamma$). For non-correlated semiconductors, the vertex contributions are  only of the order of 0.2~eV for states close to the Fermi-level but
can reach up to 1~eV for localized $d$-orbitals \cite{Gruneis_IP_solids_2014}.

With the evaluation of $\Sigma^c$, a single shot $G_0W_0$ calculation is finished, so it is worthwhile to recapitulate what can be done with the yet calculated quantities. It is straightforward to express the self energy
in any basis, for instance,  a set of localized Wannier functions  and to export it to a DMFT solver. The advantages over
a conventional $GW$ implementation are numerous. (i) First, many $GW$ codes avoid calculating the full frequency
dependency of the self energy, and instead evaluate the self energy only at a few points close to the DFT one-electron
energies $\Sigma(\epsilon^{\rm DFT}_i)$. In the present code, this is no longer necessary and one obtains the
self energy at all imaginary time points by Eq.~(\ref{eu:selfenergy}). There
is a (small) price to pay, though: to obtain physically measurable quantities, the self energy needs to be continued
to the real axis, for which continued fractions are used~\cite{Liu2016}. However, since DMFT solvers usually work
in imaginary time, the interface between VASP and DMFT is simple and requires only an interpolation from
the few available imaginary frequency points $\{\ii\omega_i\}$ to a denser Matsubara grid. (ii) Each of the individual compute steps
scales (at worst) cubic in the number of grid points or plane waves and linear in the number of k-points, as opposed to conventional
$GW$ codes, which scale quartic in the number of basis functions and quadratic in the number of k-points. The
favorable scaling  is straightforward to see: the calculation of the  polarizability  [Eq.~(\ref{eq:chi_GG})] and self energy [Eq.~(\ref{eu:selfenergy})]
are clearly quadratic in the number of grid points, however, cubically scaling rank one updates of matrices and matrix multiplications
are required to calculate the Green function [Eq.~(\ref{eq:Green_occ})] and the screened potential [Eq.~(\ref{eq:W})].
This favorable scaling combined with the efficient frequency grids allowed for efficient calculations
of the random phase approximation (RPA) of the correlation energy for isolated defects in huge supercells containing
several hundred atoms~\cite{KaltakPRB2014}.
(iii) The constrained RPA (cRPA)~\cite{Aryasetiawan2004} is simple and straightforward to  implement in the present code. One only needs to remove the
polarizability $P^t(\ii \tau)= 2 G^t(\ii\tau) G^t(-\ii\tau)$
of some target, say $t_\text{2g}$, orbitals from the total polarizability $P= P^r+P^t$ to obtain an effective screened interaction $U$:
\begin{equation}
\label{eq:cRPA}
 U^{-1}(\ii \omega)= V^{-1} - P^r(\ii \omega)\quad \Leftrightarrow \quad U^{-1}(\ii \omega)= W^{-1}(\ii \omega) + P^t(\ii \omega).
\end{equation}
The polarizability $P^r(\ii \omega)$ then captures all screening effects, except for the one inside the target space,
which will be treated in the DMFT solver. The full frequency-dependent $U(\ii \omega)$ can be calculated with very little extra cost, and
after transformation from the plane wave
basis to a localized target space, it can be directly imported into a DMFT continuous time quantum Monte Carlo solver.

 The advantages of the imaginary time and imaginary frequency representation are more
obvious if self-consistency is considered. Once the $GW$ self energy is known, the Green function can be
updated by
\begin{equation}
\label{eq:selfenergy}
	G^c(\ii\omega )  = (\ii w +\mu - H^{\rm HF} - \Sigma^c +[\Sigma_{imp} -\Sigma^{GW}_{imp}])^{-1}- (\ii w +\mu - H^{\rm HF})^{-1} ,
\end{equation}
where $H^{\rm HF}= -\nabla^2/2 + V^{\rm ion}+  V^{\rm H} + V^x$ is the Hartree-Fock Hamiltonian consisting of the kinetic energy term,
the ionic $V^{\rm ion}$, Hartree $V^{\rm H}$ and exact exchange potential $V^x$. To obtain a converging Fourier
transformation when transforming to the imaginary time, the Hartree-Fock Green function (second term) needs to be subtracted
and added back in imaginary time
\begin{equation}
 G(\ii \tau )  = G^c(\ii \tau ) + G^{\rm HF} (\ii \tau ).
\end{equation}
This closes the cycle and allows to continue with a re-evaluation of the independent particle-hole polarizability in Eq. (\ref{eq:chi_GG}).
Clearly, it is also possible to add any local self energy in Eq. (\ref{eq:selfenergy})
(terms in square brackets) and, thus, seamlessly incorporate DMFT results. Likewise, the irreducible polarization propagator
can incorporate local effects beyond the independent particle-hole approximation, if the DMFT code provides
the required information ($P \rightarrow P^{GW} + P_{imp}- P^{GW}_{imp}$, compare next section). This opens the route towards a concise implementation of {\it GW}+DMFT,
as discussed in the next section (see Fig. \ref{fig:scheme}).
A closure of the self-consistency cycle is already possible in the present code,
although some intricacies for metallic systems still need to be solved, including an approximate inclusion
of Drude-like metallic screening,  and an efficient update
of the chemical potential, which is important to achieve robust convergence in
the self-consistency cycle for metals.

\subsection{Results}

The $GW$ method has now been used for almost five decades. However, despite its undisputed improvements
compared to DFT, results vary significantly between different codes. Errors are
actually particularly large for transition metal compounds  placing a serious question mark
on any quantitative predictions. Specifically in oxides,  $d$-binding energies can vary by up to 1~eV using different codes
and implementations \cite{KlimesKaltak_predictiveGW_2014}. This is clearly
unacceptable, if one aims to merge $GW$ with more accurate methods such as DMFT.
One major problem of the  $GW$ method is that the convergence with respect to the basis set size is extremely slow \cite{shih2010,friedrich2011}. Specifically,
for the projector augmented wave method, as used in VASP, the partial waves, which are supposed to form
a sufficiently complete basis in the vicinity of the atoms, need to be chosen such that basis set convergence can be attained.
The slow convergence has been rigorously discussed by Klimes, Kaltak and Kresse in Ref. \cite{KlimesKaltak_predictiveGW_2014}.
Although that paper also establishes a suitable benchmark for solid state systems, we are not
aware that other comparable reference numbers have yet been published for solids.
Then, how can one ascertain that the numbers predicted with VASP are accurate and
reproduce the infinite basis set limit? Fortunately, the new VASP $GW$ code allows us to address this issue. Since it is efficient for
large unit cells and large basis sets, it is possible to compare the results for molecules with atomic codes
that use  Gaussian type orbitals (GTOs). GTOs have been used for 50 years in quantum
chemistry and have matured to a point where convergence for excited state calculations can
be obtained fairly easily, although careful basis set extrapolations are as important as for plane waves.

Fig. \ref{Fig3} shows the difference between the basis set extrapolated GTO results
and the VASP PAW results for the ionization potential of 100 closed shell molecules.
The mean deviation between both codes is only 60 meV \cite{MaggioPeitao_GW100_2017}, and large
outliers are practically absent. We note that the deviations between other plane wave
codes and GTOs are on average twice as large, but can even reach 200 meV on average.
The other important point is the large difference between theory and experiment  highlighting how limited the precision of $G_0W_0$ is even for
simple weakly correlated systems such as small molecules. This  clearly underlines the need to go beyond the random
phase approximation and single shot  $G_0W_0$ calculations.

  \begin{figure}
  \begin{center}   \includegraphics[width=1.0\columnwidth,clip]{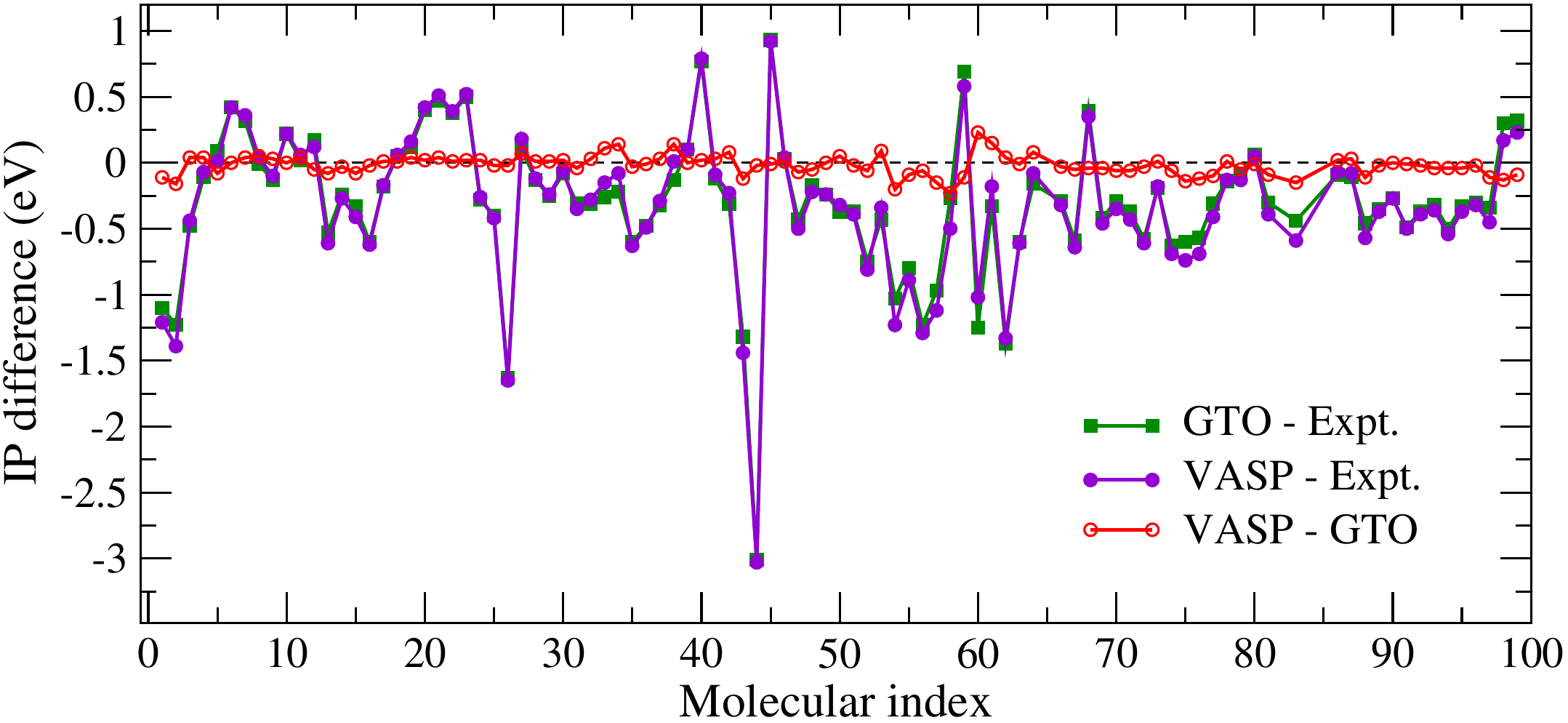}
   \end{center}
\caption{\label{Fig3}
Difference between GTO, PW and experimental values for the ionization potential (IP)
for a set of 100 molecules. The theoretical data are from $G_0W_0$ calculations using GGA orbitals
(\protect the data have been collected from Ref.~\cite{MaggioPeitao_GW100_2017}).}
  \end{figure}

\begin{figure}
\begin{center}
\includegraphics[width=1.0\textwidth]{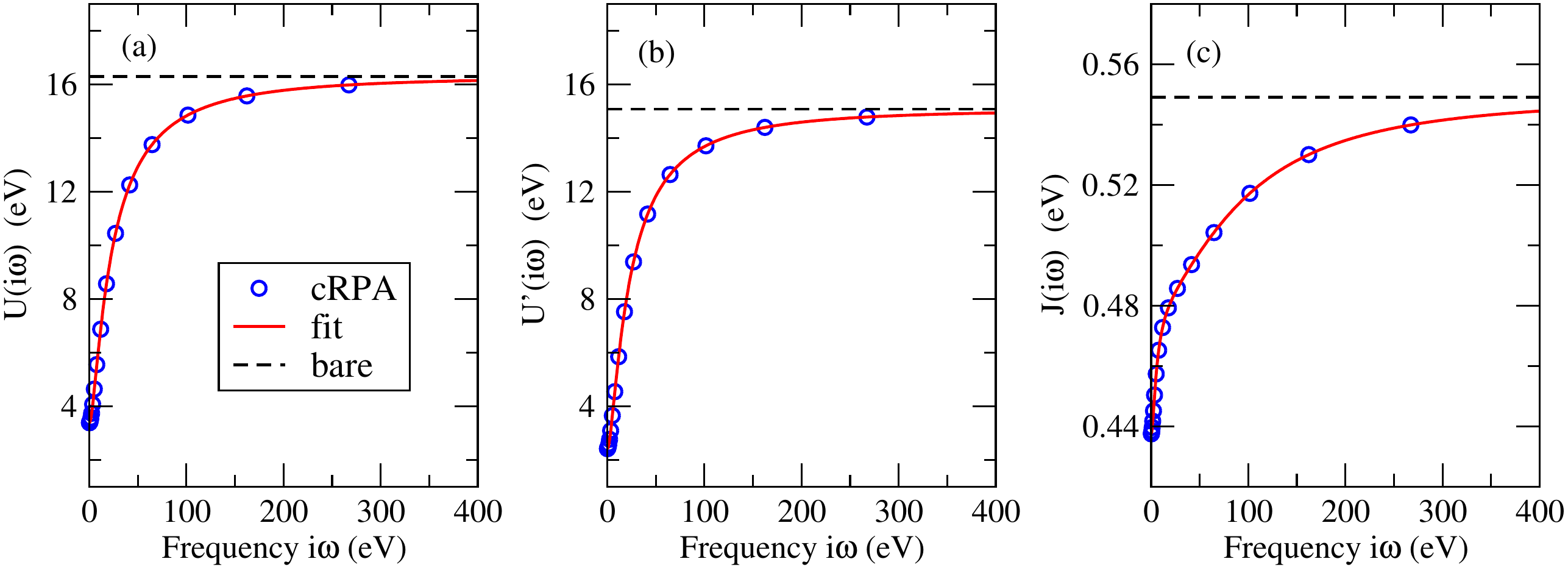}
\end{center}
\caption{ On-site dynamical partially screened (a) intra-orbital interactions $U(\ii\omega)$,
(b) inter-orbital $U'(\ii\omega)$, and (c) Hund's coupling $J(\ii\omega)$ of SrVO$_3$ as a function of
the imaginary frequency (shown in blue circle).  The bare counterparts
are also shown as black dashed lines.
The red solid lines are obtained from a Pad\'{e} fit.
We use 20 optimized imaginary frequency grid points and $8\times8\times8$ $k$ points
in the calculations.}
\label{fig:CRPAR_SrVO3}
\end{figure}

\newpage
As an illustrative example for solid state calculations, we show results for SrVO$_3$.
Fig.~\ref{fig:CRPAR_SrVO3} shows our calculated on-site dynamical screened
intra-orbital interaction $U(\ii\omega)$, inter-orbital $U'(\ii\omega)$, and Hund's coupling $J(\ii\omega)$
of SrVO$_3$ using the cRPA and V-$t_\text{2g}$-like maximally localized Wannier functions. In imaginary frequency, $U$, $U'$ and $J$ are rather smooth functions,
so that it is possible to interpolate them from the optimized frequency grid to Matsubara frequencies by a Pad\'{e} interpolation~\cite{Vidberg1977}
(see the red solid lines in Fig.~\ref{fig:CRPAR_SrVO3}). This makes it possible to transfer them to a
dynamical impurity solver.
In the static limit ($\omega=0$), $U$, $U'$, and $J$ are calculated
to be 3.38, 2.42, and 0.44 eV, respectively, agreeing perfectly with the ones
directly obtained from the conventional implementation working on
the real frequency axis. Moreover, they are in nice agreement
with the published values~\cite{Aryasetiawan06,PhysRevB.86.165105}.
In the high-frequency limit ($\omega\rightarrow\infty$), $U$, $U'$, and $J$ approach the unscreened (bare) counterparts (16.29, 15.07, and 0.55 eV).

\begin{figure*}
\begin{center}
\includegraphics[width=1.00\textwidth]{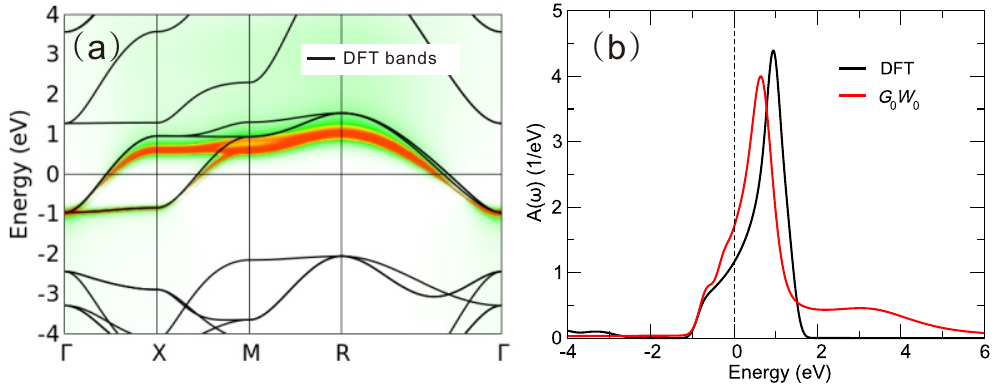}
\end{center}
\caption{
(color online)  (a)  Momentum resolved spectral function  in the  $G_0W_0$ approximation (color/gray) and DFT (black line).
 (b) Comparison of the DFT and local $G_0W_0$ spectral function $A(\omega)$.
 }
\label{fig:SrVO3_spectral}
\end{figure*}

Figure~\ref{fig:SrVO3_spectral}(a) shows the single-shot $G_0W_0$ momentum resolved spectral function.
Compared to DFT, the $t_\text{2g}$ bandwidth is reduced by 20~\% in $G_0W_0$. In
the  $G_0W_0$ approximation,  spectral weight is transferred to satellites. This is much more clearly seen in the local, momentum-integrated, spectral function as shown in Fig~\ref{fig:SrVO3_spectral}(b).
The plasmon satellite of the $t_\text{2g}$ quasi-particle band at $\sim$3 eV  arising from the $t_\text{2g}$ contribution to the fully screened interaction at the plasmon frequency~\cite{Tomczak12,PhysRevB.94.201106} is well reproduced. Further, a plasmon peak deriving from transitions outside the $t_\text{2g}$ subspace is seen at $\sim$15 eV \cite{Aryasetiawan06,Tomczak12}.
We note, however, that in our calculations repeated plasmon peaks at higher frequencies are absent. This is a well known issue of
the  $G_0W_0$ approximation \cite{PhysRevLett.107.166401}.


\section{Screened exchange and local quantum fluctuations: {\it GW}+DMFT}
\label{Sec:GWDMFT}
\subsection{Method}

The key advantage of the {\it GW} approach discussed above is its treatment of dynamical screening:
While standard electronic structure methodologies---such as Hartree-Fock or DFT---work with the
bare Coulomb interaction $V$, {\it GW} explicitly incorporates the polarizability of the electronic system.
As a consequence, the repulsion between electrons becomes reduced and retarded.
The resulting screened-exchange self energy yields a much improved description of, e.g., $sp$-semiconductors gaps, whereas the retardation effects
account for spectral weight transfers to (plasmon) satellite features, and finite lifetimes of electronic excitations.

However, as already mentioned in the Introduction,  the perturbative {\it GW} approach is insufficient for strongly correlated materials.
In fact, it fails  to account for their strong mass renormalizations, Hubbard satellites, and local moments physics \cite{Imada1998}.
Our recent understanding of strong electron-electron correlations was indeed propelled by the advent of a non-perturbative technique:
the dynamical mean field theory \cite{Georges1996}. The latter maps the lattice problem onto the self-consistent solution of an Anderson impurity model,
and the lattice self energy is identified with the single-site (i.e., local) self energy of the impurity \cite{Georges1992a}. This mapping becomes exact in the limit
of infinite lattice coordination \cite{Metzner1989}. By construction, DMFT accounts only for correlations from on-site interactions,
yet it includes---as depicted in Fig.~\ref{Fig1}---all Feynman diagrams built from the Hubbard $U$ and Hund $J$ interactions and the local impurity propagator.
To set up a realistic DMFT calculation, the one-particle part of the Hamiltonian is taken from DFT (whence the
name DFT+DMFT \cite{Lichtenstein1998,Anisimov1997}) and the screened interaction parameters---$U$ and $J$---can be computed from techniques such as constrained DFT \cite{constrainedLDA}, or, better, the constrained
random phase approximation \cite{Aryasetiawan2004,miyake:085122,jmt_mno}. However, as mentioned in the Introduction, it is not separable how the Hubbard $U$ already contributes to the DFT band-structure.
so that there is the problem of ``double-counting'' correlations when adding the DMFT self energy.

From this brief summary it is apparent that {\it GW} and DMFT are very complementary techniques:
{\it GW} has no restriction on the range of the interaction or the self energy and therefore excels for $sp$-systems.
In DMFT the interaction and the self energy are by necessity localized on an atomic site, yet their non-perturbativeness
allows for a reliable description of the Kondo and Mott physics realized in many $d$- or $f$-electron materials.
At the same time, {\it GW} and DMFT share a common (diagrammatic) language. 
Therewith, both methods can profit from each other:
As the RPA technique is integral part of the {\it GW}, it can provide the DMFT with a Hubbard $U$ computed from first principles [see the preceding section and Eq.~(\ref{eq:cRPA})].
In return, DMFT susceptibilities and self energies can add local vertex corrections to all orders to Hedin's equations for the polarization and self energy, see Eqs.~(\ref{eq:chi_GG}) and (\ref{eu:selfenergy}), respectively.
Contrary to DFT+DMFT, any double-counting in this combination of screening and correlations can be avoided, since a clear-cut separation is possible on the diagrammatic level.

This outlines the {\it GW}+DMFT method proposed in Ref.~\cite{Biermann2003}. By elegantly combining the best of both worlds---screened exchange and local quantum fluctuations---{\it GW}+DMFT has the potential
to vastly extend the realm of quantitative and predictive many-body electronic structure theory.
Let us give specific examples: In many materials the separation between correlated $d$ or $f$-states and ligand $sp$-orbitals is often severely underestimated
within DFT and DFT+DMFT \cite{PhysRevB.77.205112,0953-8984-23-8-085601,PhysRevB.93.235138,Hansmann2014}. Yet, optical transitions between these states can actually be relevant for technological applications in, e.g., intelligent window coatings \cite{optic_epl},
or eco-friendly rare-earth-based pigments \cite{jmt_cesf}. Calculating the red colour of CeSF indeed required incorporating a {\it GW} correction into DFT+DMFT \cite{jmt_cesf}.
Non-local (inter-site) self energies {\`a} la {\it GW} were also shown to be crucial in
oxides \cite{PhysRevB.89.235119,PhysRevB.93.235138}, intermetallics \cite{PhysRevB.82.085104}, and
iron-pnictides and chalcogenides \cite{jmt_pnict}, and in particular for explaining the non-magnetic nature of BaCo$_2$As$_2$ \cite{paris_sex}.
Moreover, important effects of dynamical screening were found, among others, in oxides \cite{PhysRevB.85.035115,Tomczak12,0295-5075-99-6-67003}, pnictides \cite{Werner2012,paris_sex} and cuprates \cite{PhysRevB.91.125142}.
On the other side, local vertex corrections in susceptibilities beyond RPA where shown to be crucial in, both, Hubbard models \cite{Ayral2012,Ayral2013} and realistic materials, e.g., regarding
the dynamical structure factor in iron-pnictides \cite{Park2011a,Toschi2012}, and the absence of ferromagnetism in stoichiometric FeAl \cite{PhysRevB.92.205132}.

	\begin{figure}[h!t!]
	\begin{center}
			{\includegraphics[width=1.05\columnwidth,trim={5cm 17cm 0cm 4cm},clip]{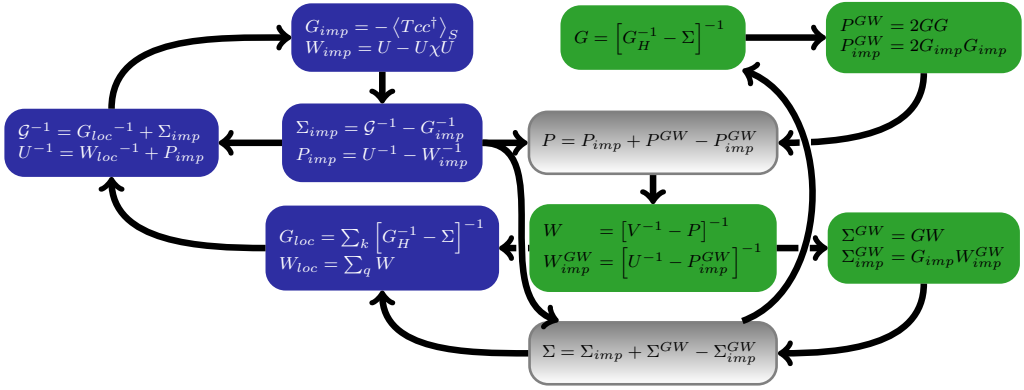}}
    \caption{\label{fig:scheme} The {\it GW}+DMFT approach. The DMFT sub-cycle is indicated in blue, the {\it GW} procedure in green, and shared quantities in grey boxes.
				}
		\end{center}
  \end{figure}

\medskip

After this general rationale, we will now discuss {\it GW}+DMFT in some more detail (see Refs.~\cite{gwdmft_proc1,Held2011,0953-8984-26-17-173202,0953-8984-28-38-383001} 
for longer reviews). The workflow of the approach is depicted in Fig.~\ref{fig:scheme}:
On the left---in blue--is the  DMFT \cite{PhysRevB.52.10295,Si1996,Sun02} cycle with an additional self consistency for the two-particle interaction:
It is required that the local screened interaction $W_{loc}$ equals the screened impurity interaction $W_{imp} =U + U P_{imp}W_{imp} = U- U \chi U $,
where $\chi=\langle \mathcal{T} n(\tau)n(0) \rangle$ is the impurity density-density correlation function and $U$ the local interaction (containing, e.g., Hubbard and Hund terms).
Owing to the dynamical nature of screening, these interactions are in particular frequency-dependent, i.e., $U\rightarrow U(\omega)$.
At least for density-density type of terms, solving an Anderson impurity model with dynamical interactions is easily possible with quantum Monte Carlo techniques, both approximately \cite{PhysRevB.85.035115} and numerically exactly  \cite{PhysRevB.85.035115,Werner10,PhysRevB.92.115123}.
On the right---in green---are Hedin's equations for the polarization $P^{GW}$ and the self energy $\Sigma^{GW}$ in the {\it GW} approximation.
Neglecting vertex-corrections here boils it down  to the RPA for $P^{GW}$, Eq.~(\ref{eq:chi_GG}), and the first order $GW$ expression for $\Sigma$, Eq.~(\ref{eu:selfenergy}).

DMFT and {\it GW} intersect at two junctures---marked in grey---once on the two-particle level in the polarization, and once on the one-particle/spectral level in the self energy.
Both times, non-perturbative, yet local contributions from DMFT, $P_{imp}$ and $\Sigma_{imp}$, are added to the {\it GW} contributions, $P^{GW}$ and $\Sigma^{GW}$.
Since the latter already contain some of the local diagrams of the former, these terms have to be subtracted.
Indeed, at each iteration, we need to remove all contributions from the polarization and the self energy that arise when computing the impurity analogues of $P$ and $\Sigma$
{\it on the {\it GW} level}.
In case of the polarization, this is achieved by subtracting a local RPA polarization $P^{GW}_{imp}=2G_{imp}G_{imp}$ obtained from a convolution of two impurity Green functions \cite{Nomura12}: $P=P_{imp}+P^{GW}-P_{imp}^{GW}$.
For the self energy, we need to subtract a term $\Sigma_{imp}^{GW}$ that is computed as the first order contribution in an interaction $W^{GW}_{imp}$ that derives from
screening the impurity interaction---the Hubbard $U$---with the above polarization $P^{GW}_{imp}$, i.e., $W^{GW}_{imp}=\left[U^{-1}-P_{imp}^{GW}\right]^{-1}$.%
\footnote{Here, we leave out details on how to connect DMFT and {\it GW} in orbital space. For this aspect see the ``orbital-separated'' {\it GW}+DMFT scheme in Ref.~\cite{Tomczak14} and also Ref.~\cite{PhysRevB.94.201106}.}
Due to these junctures, there is an outer self-consistency that allows for a feedback of local and non-local many-body effects onto the {\it GW} and DMFT cycles, respectively.
Typically such a calculation is initialized with a Green function $G=\left[G_H^{-1}-\Sigma\right]^{-1}$, where $G_H$ denotes the Hartree Green function, and a guess for the self energy $\Sigma$
(here including the Fock term).
In the first iteration, $\Sigma$ is usually replaced by the DFT exchange-correlation potential $V^{xc}$, i.e., $G=G^{DFT}$ (called $G_0$ in the Introduction section).

\subsection{Results}

Combining two methods that have evolved and matured independently over decades into large software packages is an intricate endeavor.
Therefore, the full scheme, as shown in Fig.~\ref{fig:scheme}, has been realized  first for one-band calculations \cite{Sun02,Ayral2012,Ayral2013,Hansmann2013}.
For realistic multi-band systems, the first implementation---Tomczak {\it et al.} \cite{Tomczak12}---resorts to simplifications, namely (1) omitting global self-consistency, i.e.,
performing only one-shot {\it GW} calculations starting with $G=G^{DFT}$, (2) fixing the double-counting polarization $P^{GW}_{imp}$ to the (dynamical!) cRPA result and approximating $P_{imp}\approx P^{GW}_{imp}$,
(3) approximating the double-counting self energy by the local projection of the {\it GW} self energy: $\Sigma^{GW}_{imp}\approx\sum_k\Sigma^{GW}$,
and (4) solving the DMFT impurity with dynamical $U(\omega)$ within the approximative Bose factor Ansatz \cite{PhysRevB.85.035115}.
In other early works, additional approximations were made: Taranto {\it et al.} used a static Hubbard $U$ and circumvented computing a fully frequency-dependent $\Sigma^{GW} $\cite{Taranto2013} (see also below), and Sakuma {\it et al.} combined DMFT and {\it GW} self energies from independent calculations \cite{PhysRevB.88.235110}.

Applied to the prototypical correlated metal SrVO$_3$, {\it GW}+DMFT revealed important new insights \cite{Tomczak12,Tomczak14}:
The additional ingredients---the momentum-dependent self energy $\Sigma(\mathbf{k},\omega)$ and retardation effects
in the Hubbard interaction $U(\omega)$---are found to {\it compete}.
The dynamics in the interaction describes, among others, spectral weight transfers to plasmon satellites (at $\sim 15$eV for SrVO$_3$ \cite{Aryasetiawan06}). These high energy excitations
account for an {\it additional} (i.e., beyond Hubbard-model physics) reduction of the low-energy quasi-particle weight and, correspondingly, to a {\it narrowing} of the band-width \cite{PhysRevB.85.035115,0295-5075-99-6-67003} (by a factor $Z_B\sim 0.7$ in SrVO$_3$\cite{casula_effmodel}).
Therefore, DMFT calculations that use only static interactions, have to employ a larger  Hubbard $U=4-5.5$eV \cite{Pavarini04,Nekrasov05a,PhysRevB.77.205112,0953-8984-23-8-085601} than the static limit
$U(\omega=0)\approx3.5$eV\cite{Aryasetiawan06,miyake:085122,Tomczak12} of the cRPA to account for the same mass enhancement; such a larger interaction is actually obtained  in constrained LDA \cite{Sekiyama2004}.
The non-local exchange self energy on the other hand {\it widens} the low-energy dispersion \cite{jmt_pnict,Tomczak12,Miyake13,0295-5075-108-5-57003}.
Correspondingly, effective masses of quasi-particles are reduced.
With respect to the LDA reference, effective masses are given by the ratio of the LDA and {\it GW}+DMFT group velocities:
\begin{equation}
\frac{m^{*\phantom{DA}}}{m^{LDA}}=\frac{d\epsilon_{\bf k}^{LDA}/dk}{dE_{\bf k}^{}/dk},  \qquad \frac{dE_{\bf k}}{dk}=\left.\frac{d\epsilon_{\bf k}^{LDA}/dk+\partial_k {\rm Re} \Sigma({\bf k},\omega)}{1-\partial_\omega {\rm Re}\Sigma({\bf k},\omega)}\right|_{{\bf k}={\bf k_F},\omega=0} \; .
\label{eq:}
\end{equation}
Here, the denominator is related to the quasi-particle weight $Z_{\bf k}=[1-\partial_\omega {\rm Re} \Sigma({\bf k},\omega)]^{-1}_{\omega=0}$. In DMFT, where the self energy is local,
$m^*/m^{LDA}=1/Z$ holds. In {\it GW}+DMFT, the extra term involving the momentum derivative of the self energy substantially counteracts the mass enhancement generated by the dynamical correlations \cite{jmt_pnict,Tomczak14}.
Altogether this yields a similar effective mass as in the previous DFT+DMFT calculations (that use $U>U(\omega=0)$), but
 the low-energy spectral {\it weight}  is different,
and can be measured by transport or optics.

In Fig.~\ref{fig:self} we compare Matsubara self energies for the $t_\text{2g}$ orbitals of SrVO$_3$ obtained with
our new implementation that combines the {\it GW}-code of VASP detailed in Section 1 with the w2dynamics DMFT code \cite{Parragh12,Wallerberger16}.%
\footnote{The interface between both codes has been implemented by D.\ Springer. The capability to use retarded density-density interactions in w2dynamics has been provided
by D.\ Springer and A.\ Hausoel. The framework of the Research Unit 1346 was instrumental for the success of this collaboration involving at least 3 independent research groups.}
We employ the same approximations (1)-(3) as in Ref.~\cite{Tomczak12}. However, instead of (4)
the approximative Bose-factor Ansatz \cite{PhysRevB.85.035115}, we use a numerical exact continuous-time quantum Monte Carlo
algorithm for retarded density-density interactions \cite{Werner10}.
Our reference is a standard DFT+DMFT calculation that uses a static Hubbard $U(\omega=0)$ and Hund's $J$ as provided by the cRPA (see Fig.~\ref{fig:CRPAR_SrVO3}).
From the low-energy slope of ${\rm Im}\Sigma(i\omega_n)$ we extract a quasi-particle weight $Z=0.6$. Turning on the retardation in the interaction,
i.e.\ solving DFT+DMFT with the dynamical cRPA $U(\omega)$ adds substantial renormalizations of plasmonic origin; $Z$ decreases to 0.3.
Moreover, since the dynamical interaction recovers at high frequencies the unscreened Coulomb interaction, $ {\rm Re} U(\omega\rightarrow\infty)\approx 16$eV, also the self energy $\Sigma$
lives on a much larger energy scale \cite{PhysRevB.85.035115} than in the standard, static DFT+DMFT case.
Adding the non-local {\it GW} self energy decreases effective masses, i.e.\ the ratio of $U$ over bandwidth diminishes and so does the strength of correlations:
In  our {\it GW}+DMFT the local quasi-particle weight is $Z=0.63$, which is even slightly larger than within static DFT+DMFT.
We find qualitative agreement with previous {\it GW}+DMFT results from Ref.~\cite{Tomczak14}.\footnote{%
Deviations could be explained by differences in temperature, the lattice constant, as well as the lifting of approximation (4).}

\begin{figure}
\begin{center}
\includegraphics[width=.7\columnwidth]{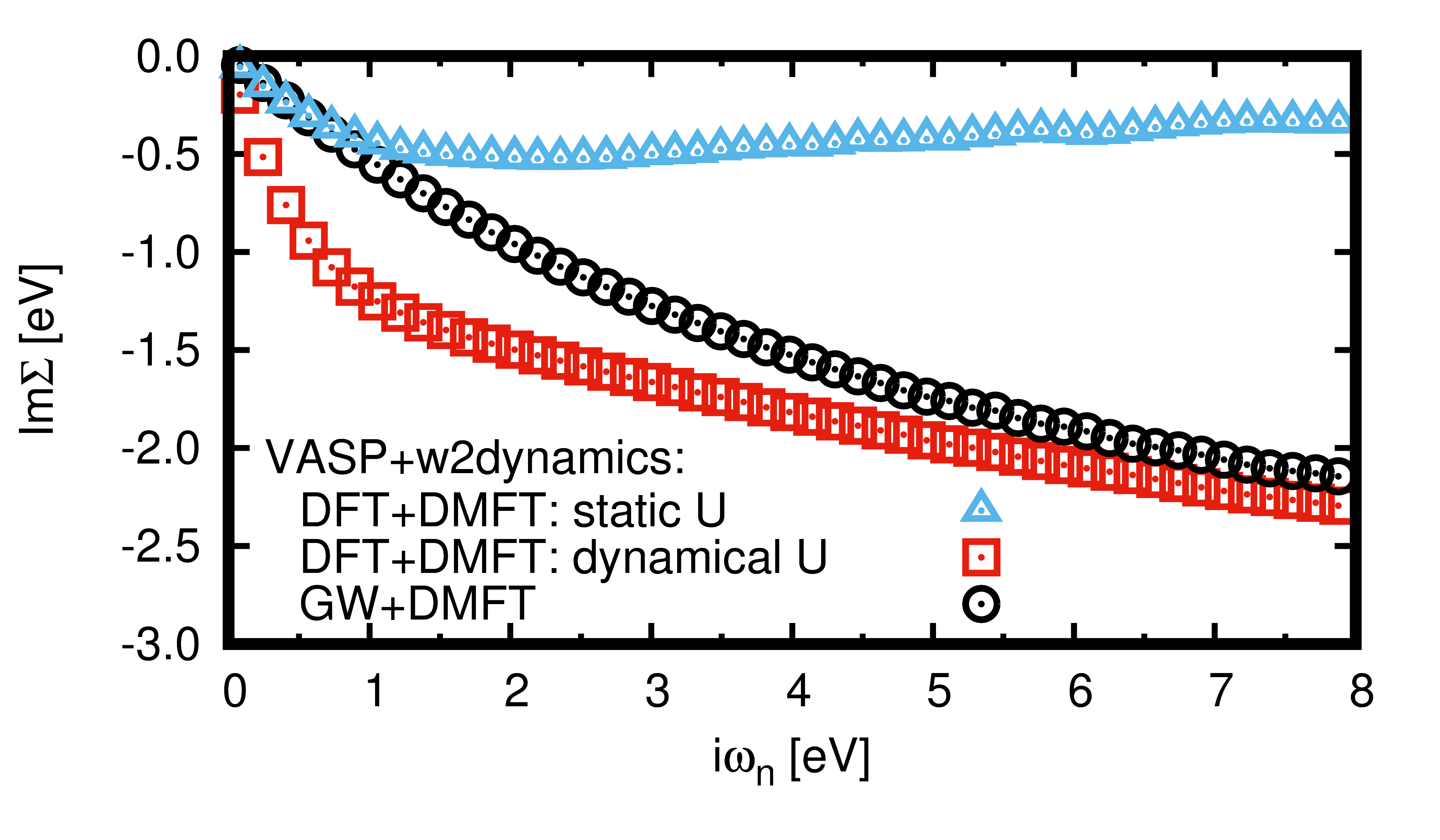}
\end{center}
\caption{Local Matsubara self energies ($T=300$K) for the $t_\text{2g}$ orbitals of SrVO$_3$ using (i) DFT+DMFT with static $U=U(\omega=0)$, (ii) DFT+DMFT with the dynamical interaction $U(\omega)$ from cRPA, and (iii) {\it GW}+DMFT.
Data obtained using the new {\it GW} implementation of VASP in combination with w2dynamics.}
\label{fig:self}%
\end{figure}

{\it GW}+DMFT spectra for the $t_\text{2g}$-orbitals of SrVO$_3$ from Ref.~\cite{Tomczak12} are shown in Fig.~\ref{fig:arpes} in comparison with angle-resolved photoemission spectroscopy (ARPES) results.
The calculation agrees well with the experimental data. Differences to previous DFT+DMFT calculations (see, e.g., Refs.~\cite{Pavarini04,Nekrasov05a,PhysRevB.77.205112,0953-8984-23-8-085601})
are however most pronounced for unoccupied states~\cite{Tomczak14}, that are inaccessible to ARPES experiments. The effects are in line with the above discussion: (i) W.r.t.\ DFT+DMFT the low-energy bandwidth is enhanced
by non-local self energy contributions (e.g., the unoccupied $d_{xy}$, $d_{xz}$-bands at the X point move up from 0.6eV\cite{Nekrasov05a} to $\sim 1$eV). (ii) Using an {\it ab initio} screened interaction $U(\omega)$ (instead of a larger static $U$ adjusted so as to reproduce the experimental mass enhancement),
the upper Hubbard band is placed at much lower energy (e.g., 1.2(1.9)eV instead of 2.2(2.85)eV \cite{Nekrasov05a} at the $\Gamma$(X) point for the $d_{xy}$, $d_{xz}$-components). Indeed, the upper Hubbard band merges with the quasi-particle peak in momentum-integrated spectra.
This reduced importance of Hubbard physics
has recently been confirmed by partially self-consistent {\it GW}+DMFT calculations \cite{PhysRevB.94.201106}, and is compatible with recent
inverse ARPES experiment.\footnote{T.\ Yoshida and A.\ Fujimori, private communication.}
Besides the shown low-energy dispersion, also the position of ligand states, in particular the O-$2p$ and Sr-$4d$ improve substantially in {\it GW}+DMFT.
Since the discussed {\it GW}+DMFT results are not globally self-consistent, the ligand states are at the same position as in $G_0W_0$ calculations \cite{Tomczak12,Taranto2013,Tomczak14,Kazuma16}.

\begin{figure}
\begin{center}
\includegraphics[width=0.7\columnwidth]{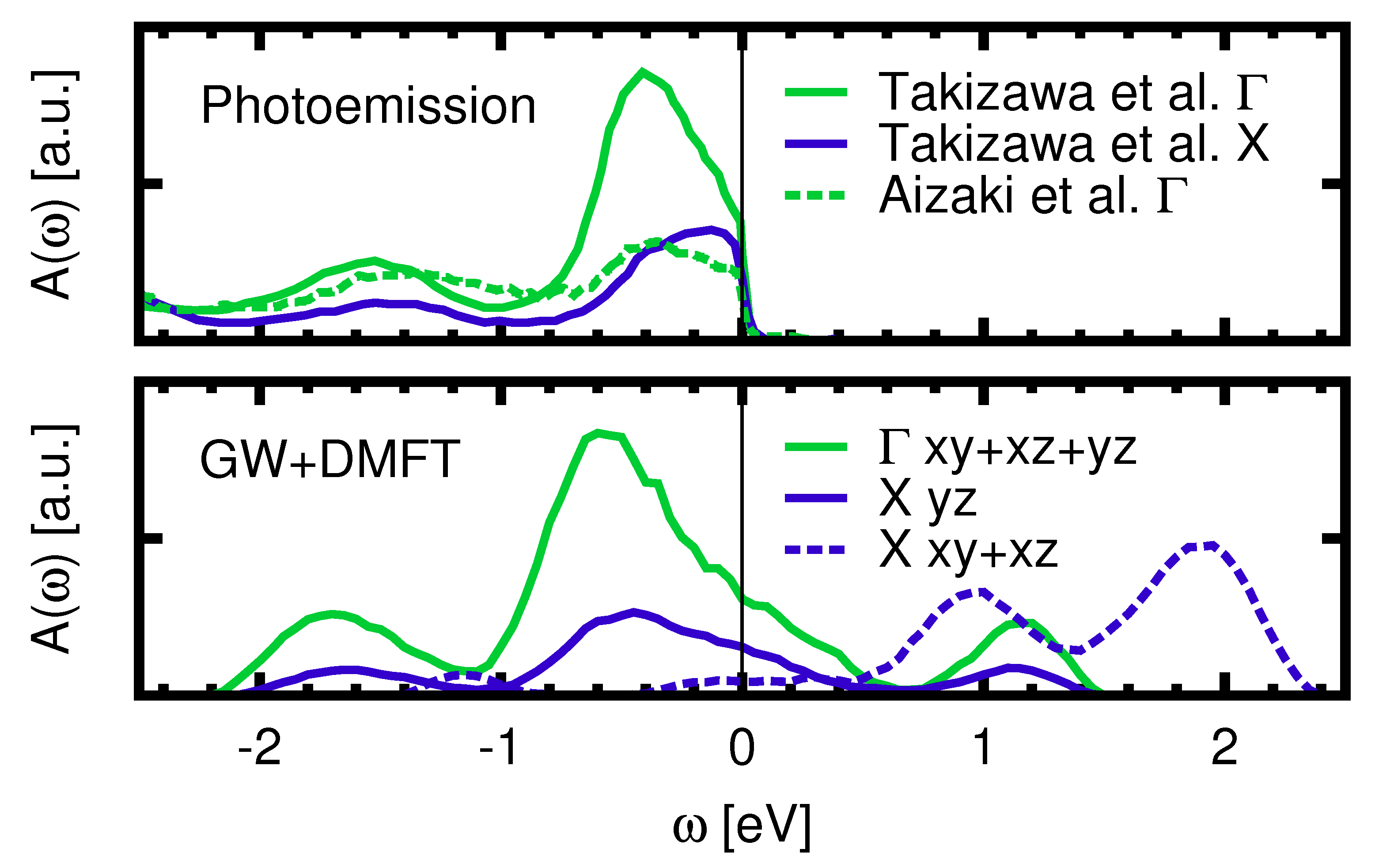}
\end{center}
\caption{Comparison of {\it GW}+DMFT spectra from Ref.~\cite{Tomczak12} (bottom) with angle resolved experimental photoemission spectra (top) from Refs.~\cite{Aizaki12,PhysRevB.80.235104} for the $\Gamma$ and $X$-point.}
\label{fig:arpes}%
\end{figure}

\medskip

In the setup used for these results (one-shot $G_0W_0$), the only modification of the DMFT code comes from adding the frequency- and momentum dependent {\it GW} self energy contributions into
the DMFT one-particle self-consistency. Yet, since the {\it GW} self energy is a large and unhandy object, it stands to reason to approximate it to alleviate memory consumption.
Also, as mentioned in the preceding section, some {\it GW} codes cannot provide the self energy on a continuous frequency mesh.
A strategy to simplify the influence of the {\it GW} self energy was pioneered by Taranto {\it et al.} in Ref.~\cite{Taranto2013}: By evaluating the self energy at the Kohn-Sham energies and hermitianizing it along the lines of Ref.~\cite{Faleev2004},
it is possible (using an approximate double-counting correction) to cast $\Sigma^{GW}$ into Hamiltonian form.
This idea was further developed into the more rigorous quasi-particle self-consistent (QS){\it GW}+DMFT approach proposed by Tomczak in Ref.~\cite{jmt_sces14}---previously alluded to in Ref.~\cite{jmt_pnict}---in which
the double-counting correction can be performed exactly, and furthermore, a global self-consistency is performed on the QS{\it GW}\cite{Faleev2004} level.
Flavors of QS{\it GW}+DMFT have subsequently been applied to cuprates and nickel oxide \cite{Choi2016}, nickel and iron \cite{PhysRevB.95.041112}, as well as insightful model systems \cite{0953-8984-27-31-315603,2016arXiv161107090L}.
Recently, Boehnke {\it et al.}\cite{PhysRevB.94.201106} have pioneered a setup in which {\it GW}+DMFT self-consistency is performed beyond the quasi-particle approximation, yet only within a low-energy subspace, in this case the $t_\text{2g}$-orbitals of SrVO$_3$.
As anticipated in earlier model calculations for the extended Hubbard model \cite{Ayral2012,Ayral2013}, the self-consistent local interactions are smaller in this setup than the initial cRPA values:
compared to previous non-self-consistent works \cite{Tomczak12,Tomczak14} the strength of correlations is reduced and mass renormalizations in SrVO$_3$ become more plasmonic in origin~\cite{PhysRevB.94.201106}.

\medskip

This concludes the description of the current state-of-the-art in {\it GW}+DMFT calculations.
While there is a panoply of materials to which the described methodology can be applied with great benefit, let us point out two challenges for future developments:
(i) Besides influencing the low-energy dispersion, the {\it GW} self energy also effects higher lying states, e.g., the O-$2p$ and the Sr-$4d$ orbitals in SrVO$_3$.
Indeed, the O-$2p$ orbitals are off by $1.5-2$eV within DFT, and are pushed towards their experimental position by the {\it GW} self energy \cite{Tomczak12,Tomczak14,Kazuma16}. This will reduce their contribution to screening, causing an {\it increase} in the Hubbard $U$ for the $t_\text{2g}$-orbitals, as indeed found when performing cRPA on top of QS{\it GW}. %
 This effect of ligand states will counteract the reduction of correlations seen in calculations in which self-consistency is limited to the $t_\text{2g}$-orbitals \cite{PhysRevB.94.201106}.
Hence, a {\it GW}+DMFT implementation that includes ligand states in the self-consistency is eagerly awaited.
(ii) Contrary to one-shot $G_0W_0$, self-consistent {\it GW}+DMFT is a conserving theory on the one-particle level:
the {\it GW}+DMFT self energy is derivable from an approximation to a free-energy functional \cite{Almbladh1999,Biermann2003}.
Yet, on the two-particle level the situation is inverted: RPA and also QS{\it GW}+DMFT~\cite{jmt_sces14} yield a conserving density-density response function/polarization.
In fully self-consistent {\it GW}+DMFT which uses dressed Green functions, on the other hand, gauge invariance for the polarization is not given and its violation
can lead to a qualitatively wrong description of, e.g., collective (plasmon) modes \cite{Hafermann2014a}.
Achieving gauge-invariance respecting one- {\it and} two-particle quantities within a {\it GW}+DMFT scheme remains a challenge for future works.


\section{Correlations on all time- and length-scales: {\em Ab initio} D$\Gamma$A}
\label{Sec:DGA}
\subsection{Method}

The essential approximation of DMFT is that the self energy $\Sigma$,
which is nothing but the  one-particle fully irreducible vertex, is local---given by all local skeleton diagrams \cite{Georges1996}. We can put this concept on the next level, assuming the locality of the two-particle fully irreducible vertex $\Lambda$.\footnote{Defined as all Feynman diagrams with two incoming and outgoing particle lines that cannot be separated into two pieces by cutting two Green function lines.} This is the  dynamical vertex approximation (D$\Gamma$A) \cite{Toschi07,Katanin2009}. From the local, fully irreducible vertex  $\Lambda$, extracted by inverting the local parquet equation of DMFT\cite{Rohringer2012}, one can construct---through the self-consistent solution of the parquet equation of the lattice system---the non-local full vertex $F$, and from that, the non-local D$\Gamma$A self energy  as well as all physical susceptibilities. This full-fledged parquet D$\Gamma$A approach has been employed in Refs.~\cite{Valli2015,Li2016}. In most calculations however, a restriction to the particle-hole (and transversal particle-hole channel) has been employed. In this so-called ladder D$\Gamma$A \cite{Toschi07,Katanin2009}, the local vertices irreducible in the two particle-hole channels $\Gamma_{ph}$ are the starting point and the full vertex $F$ is constructed through the Bethe-Salpeter ladder. This neglects the particle-particle channel which is important, e.g., for superconductivity and weak localization corrections to the conductivity.
In both variants the local and non-local self energy is obtained through the Schwinger-Dyson equation of motion, and includes non-local correlation effects such as spin fluctuations and pseudogap physics.

A variety of closely related approaches have been subsequently proposed \cite{Rubtsov2008,Rohringer2013,Taranto2014,Ayral2015,Li2015}.
They all have in common that they include all the local DMFT correlations, and
construct additional non-local correlations from the two-particle vertex
via Feynman diagrams. The differences are in the details: (i) which two-particle vertex is taken, (ii) whether the real or a dual Green function (subtracting the local Green function) is taken as connection line, (iii) which Feynman diagrams are considered. These diagrammatic extensions of DMFT have been highly successful for studying model systems such as the one-band Hubbard model and we discuss
selected results in {\bf Section \ref{Sec:HM}}.

For realistic materials calculations, one might envisage using
 D$\Gamma$A instead of DMFT in a DFT+D$\Gamma$A scheme. However, it is more appealing to use the Bethe-Salpeter equation also as a means for calculating the non-local exchange and correlation. This is possible by taking, as the irreducible vertex in the particle-hole channel, the non-local Coulomb interaction $V^q$ in addition to the local vertex, see Fig.\ \ref{Fig:DGA} (b).
Besides non-local interactions, such a  treatment also allows us to include less strongly correlated orbitals---without the need
to  calculate the local vertex for them. In the following we discuss this  AbinitioD$\Gamma$A, while results for SrVO$_3$ are presented in
 {\bf Section \ref{Sec:SVO}}

  \begin{figure}[tb]
 \begin{center}
\includegraphics[width=1.25\columnwidth,trim={1.9cm 0 0 0}]{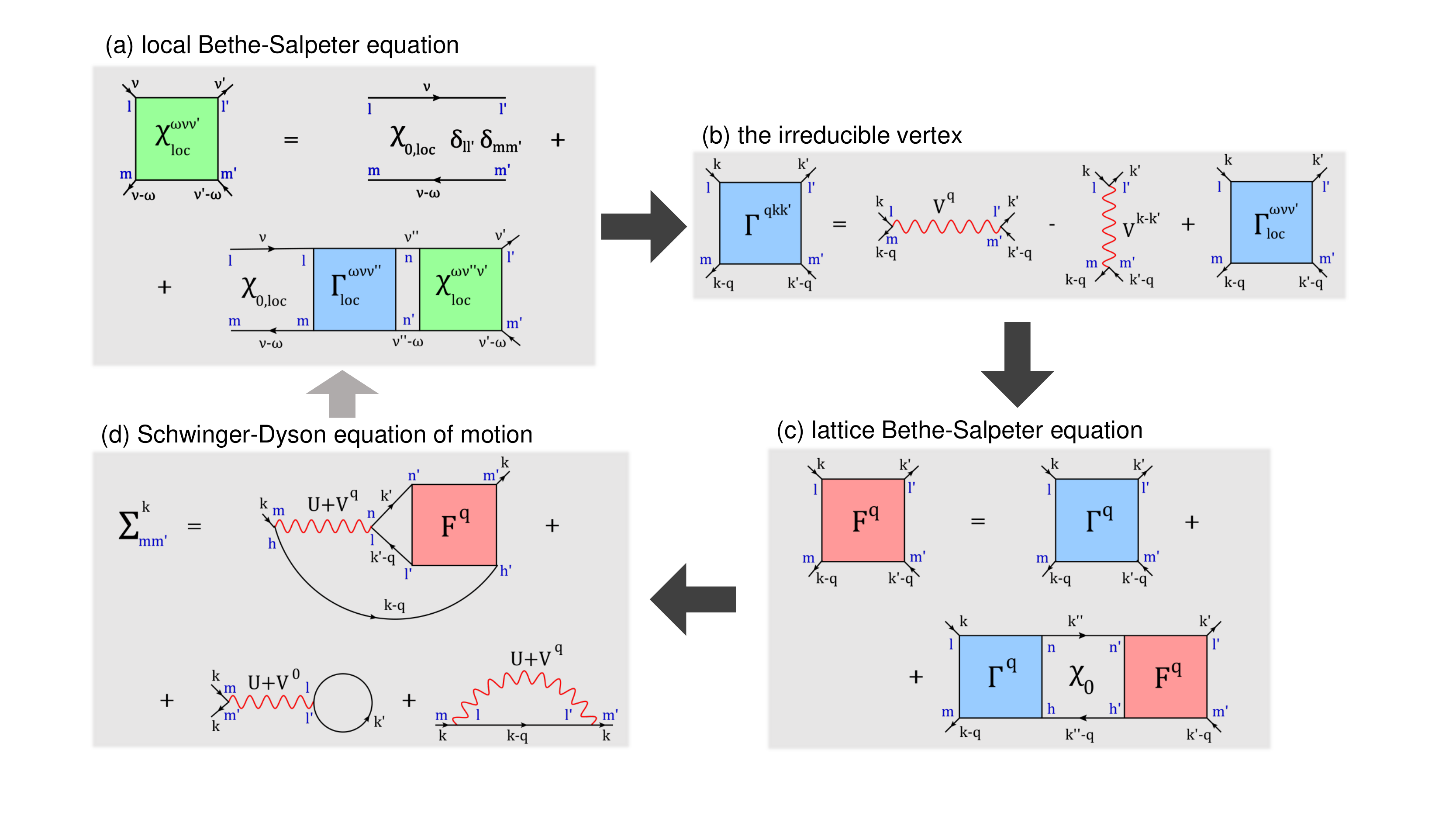}
    \end{center}
    \caption{\label{Fig:DGA} Flow diagram of the AbinitioD$\Gamma$A algorithm (from Ref.~\cite{Galler2016}). }
  \end{figure}

The flow diagram of the AbinitioD$\Gamma$A algorithm is given in Fig.\ \ref{Fig:DGA}, for a complete presentation and technical details see Ref.~\cite{Galler2016}:

{\bf Fig. \ref{Fig:DGA} (a):} The first step is to calculate the local generalized susceptibility $\chi_{\rm loc}$ via the numerical solution of an Anderson impurity model,\footnote{Calculating this vertex by continuous time quantum Monte Carlo simulations \cite{Gull2011a} is computationally the most demanding step. For getting all components of the vertex, a worm sampling is needed \cite{Gunacker15}; using an improved estimator \cite{Hafermann2014,Gunacker2016} and vertex asymptotics \cite{Kunes2011,Li2016,Wentzell2016,Gunacker2016} increase the accuracy and size of the frequency box.} and use  the  local variant of the Bethe-Salpeter equation as well as the bare (bubble) susceptibility  $\chi_{\rm 0,loc}$ to extract the local irreducible vertex in the particle-hole channel $\Gamma_{\rm loc}$ (as indicated). This vertex and the local susceptibility depends on three frequencies $\nu$, $\nu'$ and $\omega$, and four orbitals $m$, $l$,  $m'$, $l'$.

{\bf  Fig. \ref{Fig:DGA} (b):} We supplement this local irreducible vertex  $\Gamma_{\rm loc}$
with the non-local Coulomb interaction $V^q$ at momentum $q$. Together these terms  form the  AbinitioD$\Gamma$A approximation for the irreducible vertex $\Gamma^q$ in the particle-hole channel.

{\bf  Fig. \ref{Fig:DGA} (c):}  With this $\Gamma^q$ we solve the  Bethe-Salpeter equation on the lattice to get the full vertex $F^q$.\footnote{As detailed in Ref.~\cite{Galler2016}, besides the displayed particle-hole ladder, also the transversal particle-hole ladder is taken into account, and the double-counted contribution is subtracted. The Bethe-Salpeter equation is formulated in terms of a magnetic and density combination of spins which are not displayed in  Fig. \ref{Fig:DGA}. Neglecting the second, $V^{k-k'}$ term in  Fig. \ref{Fig:DGA} (b) simplifies the momentum dependence ($\Gamma^{q k k'}\rightarrow \Gamma^q$) and dramatically reduces the computational effort to solve  the Bethe-Salpeter equation.  Please note that a corresponding local contribution $U$ is included as part of  $\Gamma_{\rm loc}$ but does not lead to a $k$, $k'$-dependence.}

{\bf Fig. \ref{Fig:DGA}  (d):} This $F^q$ allows us in turn to calculate the AbinitioD$\Gamma$A  self energy
via the Schwinger-Dyson equation of motion (the second line represents the Hartree and Fock contribution to the self energy).

{\bf Self consistency:}
With a new self energy and local Green function we can, in principle,
go back to Fig. \ref{Fig:DGA}  (a) and recalculate the local susceptibility and vertex, closing the self-consistency loop.

Before turning to our presentation of selected D$\Gamma$A results,  let us briefly discuss what kind of physics the AbinitioD$\Gamma$A can describe.
First of all, we notice that the first, $V^ q$ term of  Fig. \ref{Fig:DGA} (b)
yields the RPA screening when inserted into the Bethe-Salpeter equation in the particle hole of Fig. \ref{Fig:DGA}  (c). Via Fig. \ref{Fig:DGA}  (d) this yields the $GW$ self energy. That is, all $GW$ diagrams are included in  AbinitioD$\Gamma$A.
But on top of {\it GW}, there is also the crossing symmetrically related ladder in the transversal particle-hole channel. Second, if we only consider the local Green functions in the Bethe-Salpeter equation  Fig. \ref{Fig:DGA}  (c), we recover the local $F$ or susceptibility $\chi$ of  Fig. \ref{Fig:DGA}  (a) as well as, via the equation of motion, i.e., the DMFT self energy. In other words, all DMFT diagrammatic contributions are also included.
Beyond both, we have more diagrams and physics included, e.g., spin fluctuations. These can be described in weak coupling perturbation theory as the particle-hole and transversal particle-hole ladder with the local bare interaction $U$ as
a building block. These diagrams are generated in Fig. \ref{Fig:DGA}  (c) when taking the bare $U$ term which is part of $\Gamma_{\rm loc}$ in Fig. \ref{Fig:DGA}  (b). More precisely it is the $U$ contribution to the vertex which is analogous to the second, $V^{k-k'}$  term on the right hand side of  Fig. \ref{Fig:DGA}  (c) which generates the spin fluctuations. Let us emphasize that in D$\Gamma$A such spin fluctuations are not restricted to weak coupling, the same kind of diagrams are also generated with the full  $\Gamma_{\rm loc}$.


\subsection{Results I: One-band Hubbard model}
\label{Sec:HM}

In the last decade the D$\Gamma$A has been intensively applied to study
 several physical aspects of the single-orbital Hubbard model.
On the one hand, this was important to demonstrate the performance of
D$\Gamma$A-based algorithms to describe intermediate-to-strong-coupling parameter
regions, hardly accessible to other techniques, in view of
subsequent applications to realistic systems. On the other hand, the
D$\Gamma$A, since its first applications to single-orbital models, has allowed significant progress in the fundamental understanding of important
topics in many-body physics. We just mention here, among others, in $d=2$ the
transformation of  the Mott metal-insulator transition into a crossover
down to $U=0$ and the spin-fluctuation-driven pseudogap \cite{Katanin2009,Schaefer2015-2,Schaefer2015-3,Rohringer2016a}, and, in $d=3$, the critical exponents of the Hubbard model and the breakdown of the paramagnetic Fermi-liquid
at low temperatures ($T$) because of spin fluctuations \cite{Rohringer2011,Rohringer2016a}. Notably, several of these
D$\Gamma$A findings have been supported \cite{Otsuki2014,Hirschmeier2015}  by complementary results of other powerful diagrammatic-extensions of DMFT, such as
the dual-fermion \cite{Rubtsov2008} and dual-boson approaches \cite{Rubtsov12},
as well as other novel
many-body techniques (e.g., the fluctuation diagnostics \cite{Gunnarsson2015}).

\begin{figure}
 \begin{center}   \includegraphics[width=0.75\columnwidth,clip]{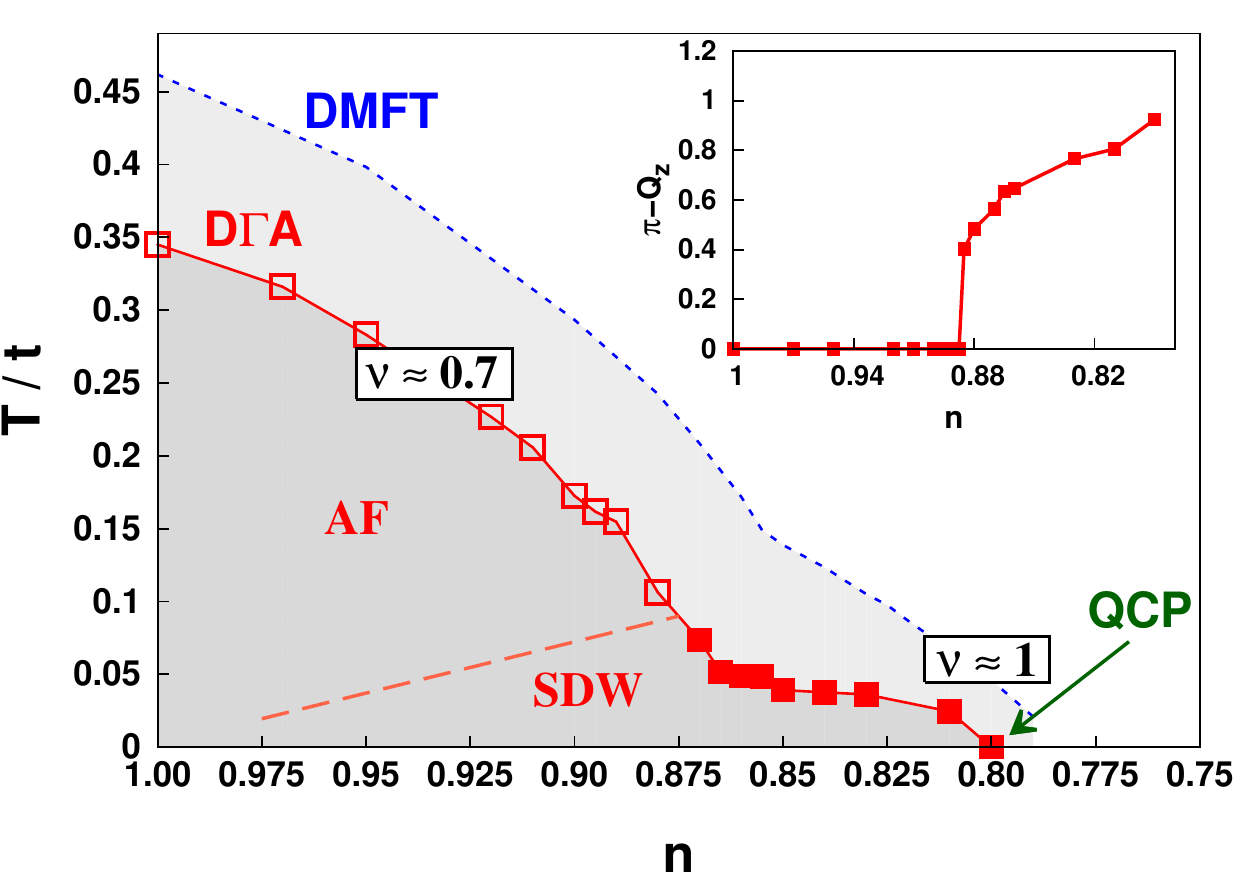}
    \end{center}
    \caption{\label{FigQCP} Magnetic phase-diagram of the three-dimensional
      Hubbard model with nearest neighbor hopping $t$ as a
      function of decreasing density $n$ (hole-doping), for an intermediate value of
      the interaction $U = 9.80 t$. The left and right box give the critical and quantum critical exponent $\nu$ at finite and zero $T$, respectively. The inset shows the deviation from commensurate AF order in D$\Gamma$A (adapted from Ref.~\cite{Schaefer2016}).
			}
  \end{figure}

In this section, we will review some of the most recent D$\Gamma$A
applications to the single-orbital Hubbard model in $d=3$, and discuss
their possible implications for the development of high-performing
multiorbital algorithms.
The first D$\Gamma$A application we consider is the investigation \cite{Schaefer2016} of the
quantum critical properties of the magnetic transition in the three
dimensional Hubbard model, as a function of (hole-)doping. As
it was also found in DMFT \cite{Jarrell1997}, the relatively high
antiferromagnetic (AF) ordering
temperature ($T_N$) of the half-filled system is progressively reduced by
increasing doping, until at about $20\%$-doping a quantum critical
point (QCP)  is found (see Fig.~\ref{FigQCP}). The reduction of $T_N$ is associated also to a gradual
transformation of the magnetic order from commensurate AF  at ($\bf \pi,\pi,\pi$)  to an incommensurate spin-density wave (SDW) at
$(\bf \pi, \pi, Q_z<\pi)$,
see inset of Fig.~\ref{FigQCP}. While the
DMFT-description of the related (quantum) critical properties is
restricted to mean-field correlations
in space, the D$\Gamma$A-treatment of both space and temporal
correlations on an equal footing yields an improved understanding of the magnetic QCPs in $d=3$. In particular,
beyond a sizable reduction of $T_N$ w.r.t.\ DMFT throughout the phase-diagram,
the finite-$T$ critical exponents $\gamma$, $\nu$ found in D$\Gamma$A  for the magnetic
susceptibility and correlation length, respectively, are
consistent with the $3d$-Heisenberg universality class (i.e., $\gamma
\simeq 1.4, \nu \simeq 0.7$), independently on whether the antiferromagnetism is
commensurate or incommensurate. While this already corrects the
mere mean-field values of critical exponents found in DMFT
($\gamma=1, \nu=0.5$),
the nature of the criticality changes further at the QCP. Here around $n\sim0.8$, the exponents take unexpectedly the
values $\gamma\simeq 0.7\div 0.8$, $\nu=1$, strongly violating the
typical scaling relation $\gamma =2 \nu$. These values are also
incompatible with the standard Hertz--Millis-Moriya theory \cite{Hertz1976,Millis1993} for perturbative QCPs. By means of a complementary semi-analytical
analysis \cite{Schaefer2016c}, the unusual values of the critical
exponents have been ascribed to the presence of lines of Kohn's points in the
underlying Fermi surface, whose effect is no longer damped by
finite-temperature fluctuations at the QCP. The D$\Gamma$A
results have, thus, identified an additional, important factor controlling the
quantum critical properties of correlated systems, hitherto mostly neglected.

The effects of non-local fluctuations are, obviously, not
confined to the (quantum) critical properties, as they also
 affect the spectral properties, in particular at the Fermi-energy.
However,  while the
electronic self energy is significantly corrected
w.r.t.\ the DMFT results, especially at low-$T$ \cite{Rohringer2011,Rohringer2016a}, a closer
inspection\cite{Schaefer2015} reveals that, in $d=3$,  the intrinsic frequency/momentum
structure of the electronic self energy in D$\Gamma$A displays specific, important
patterns. These, in turn, can be used for devising important
simplifications of realistic many-body algorithms for bulk
systems, such as {\it GW}, {\it GW}+DMFT, or the AbinitioD$\Gamma$A.
\begin{figure}
 \begin{center}
 \hspace{50mm} \includegraphics[width=1.30\columnwidth]{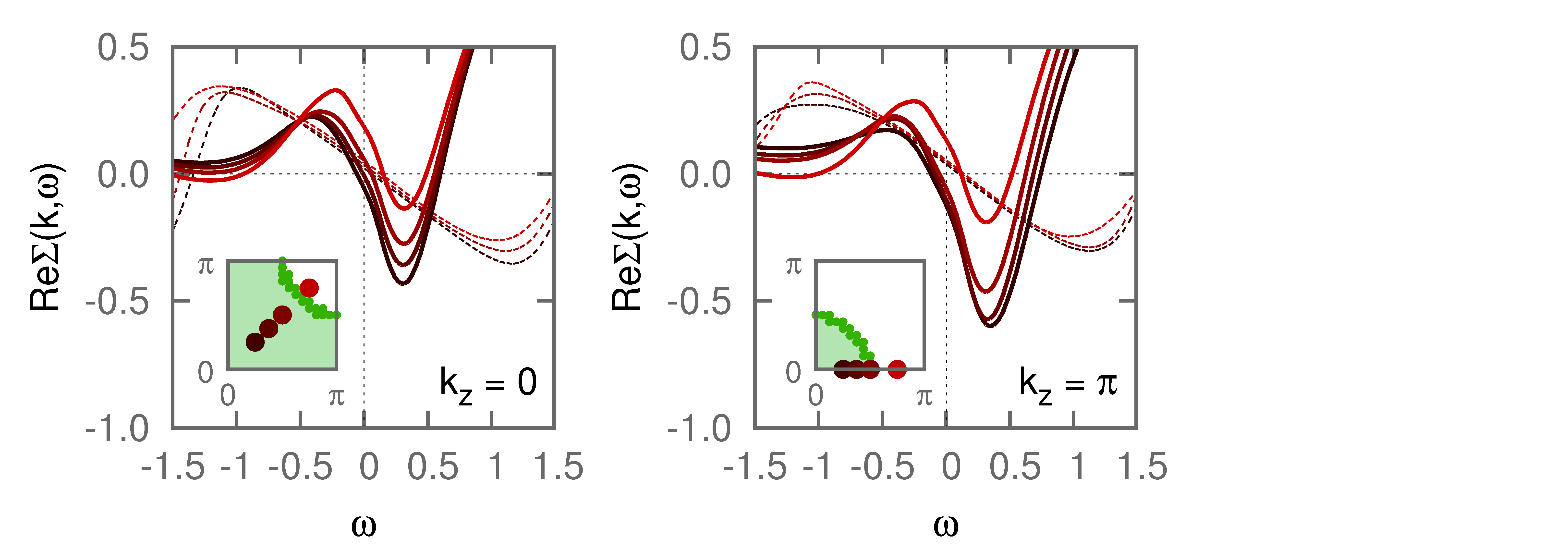}
\end{center}
 \caption{\label{FigSEPAR} Real part of the self energy $\Sigma({\bf
     k}, \omega)$ of the Hubbard model in $d=3$ ($U=1.6$, $T=0.043$ and $n=0.9$; energies in units of the half-bandwidth), analytically continued to the real-frequency axis, computed with D$\Gamma$A
   (solid lines) and $GW$ (dashed lines), respectively. The
   curves correspond to different {\bf k}-points crossing the
   Fermi-surface selected on two different paths in the $(k_x,k_y)$-planes with constant $k_z =0,
   {\pi}$ (left/right panel), as shown in the corresponding insets (adapted from Ref.~\cite{Schaefer2015}).}
  \end{figure}
Specifically, an inspection of the D$\Gamma$A self energy, continued to
real frequencies (see Fig.~\ref{FigSEPAR}) shows that the self energy of
the $3d$ Hubbard model, even in the most correlated low-doping regime ($n=0.9$),
displays a clear separation in the time/frequency and space/momentum domains:
\begin{equation}
\Sigma(\bf k,\omega) = \Sigma^{\rm loc}(\omega) +\Sigma^{\rm
  non-loc}(\bf k).
\label{eq:separ}
\end{equation}
The hallmark of such separation, which extends to a relatively broad frequency
interval around the Fermi level, is immediately visible in Fig.~\ref{FigSEPAR}: in form of the parallel frequency behavior of
Re$\Sigma(\bf{k}, \omega)$ for different ${\bf k}$, which reflects a {\sl
  momentum-independent} quasi-particle renormalization factor $Z_{\bf
  k} \sim Z$.
The D$\Gamma$A demonstrates, in fact, that the momentum dependence of $\Sigma(\bf k,\omega)$ is essentially confined to
the static sector, which explains the shift among the different
parallel self energies. Not surprisingly, the same qualitative behavior,
though---quantitatively---less
correlated (i.e., with a larger $Z$) is found in the corresponding
$GW$ results, shown for comparison in Fig.~\ref{FigSEPAR}.
It is however noteworthy that the static momentum-dependence is much larger in D$\Gamma$A than in  {\it GW}.
This advocates the presence of true non-local {\it correlation} effects as opposed to {\it exchange} effects that cause
a large (static) $\mathbf{k}$-dependence in multi-band {\it GW} calculations \cite{Miyake13,Tomczak14}.
In fact, lacking spin-fluctuations (both local and non-local), {\it GW}---where space-time separation was first evidenced\cite{jmt_pnict,Tomczak14}---verifies
Eq.~(\ref{eq:separ}) up to larger frequencies than D$\Gamma$A.

The validity of Eq.~(\ref{eq:separ}) for  strongly
correlated systems in $d\!=\!3$ is inspiring for promising algorithmic improvements,
potentially applicable to several many-body techniques (e.g., {\it GW},
{\it GW}+DMFT, AbinitioD$\Gamma$A). In all these cases, the assumption
of a full time-space separability of $\Sigma(\bf{k}, \omega)$ would
allow to avoid numerically expensive transfer-momenta/frequency
convolutions in the intermediate steps of many-body/diagrammatic calculations,
reducing considerably the numerical effort. Further details about
such simplifications, and an explicit proposal of a ``space-time separated {\it GW}'' scheme are reported in
Ref.~\cite{Schaefer2015}. We should also notice, at the end of
this section, that complementary simplifications of the momentum
structure, were suggested by the recent findings of
Ref.~\cite{Pudleiner2016}: In $d=2$ the
momentum dependence of $\Sigma(\bf{k}, \omega)$ can often be approximated by
a dependence on the non-interacting dispersion, i.e., $\Sigma(\bf{k}, \omega)
\rightarrow \Sigma(\epsilon_{\bf k}, \omega)$.

\subsection{Results II: SrVO$_3$}
\label{Sec:SVO}

Being a diagrammatic extension of DMFT, the algorithmic
implementation of D$\Gamma$A is not affected by cluster-size limitations of
cluster extensions of DMFT, making possible a systematic generalization
of the D$\Gamma$A approach to treat realistic multiorbital
systems. While the technical aspects of the AbinitioD$\Gamma$A \cite{Toschi2011,Galler2016} have been addressed in the
previous Section, here we will discuss the physics emerging from the
first applications of the AbinitioD$\Gamma$A to realistic material
calculations. Specifically, we will focus on the very recent AbinitioD$\Gamma$A study of Galler {\it et al.}\cite{Galler2016} performed for the correlated-metal testbed material SrVO$_3$.
In this compound the $3d\!-\!t_{2g}$ bands of V are rather well
separated from the other bands, which allows for a relatively accurate
modelization already in terms of  a three-orbital $t_\text{2g}$-only
manifold.
In fact, this modelization has been exploited  in the past for a huge
number of many-body calculation, from LDA+DMFT to {\it GW}+DMFT (see Section \ref{Sec:GWDMFT}), and
represents, thus, a sort of {\sl drosophila} among correlated
systems.

\begin{figure}
 \begin{flushleft}   \includegraphics[width=1.09\columnwidth,clip]{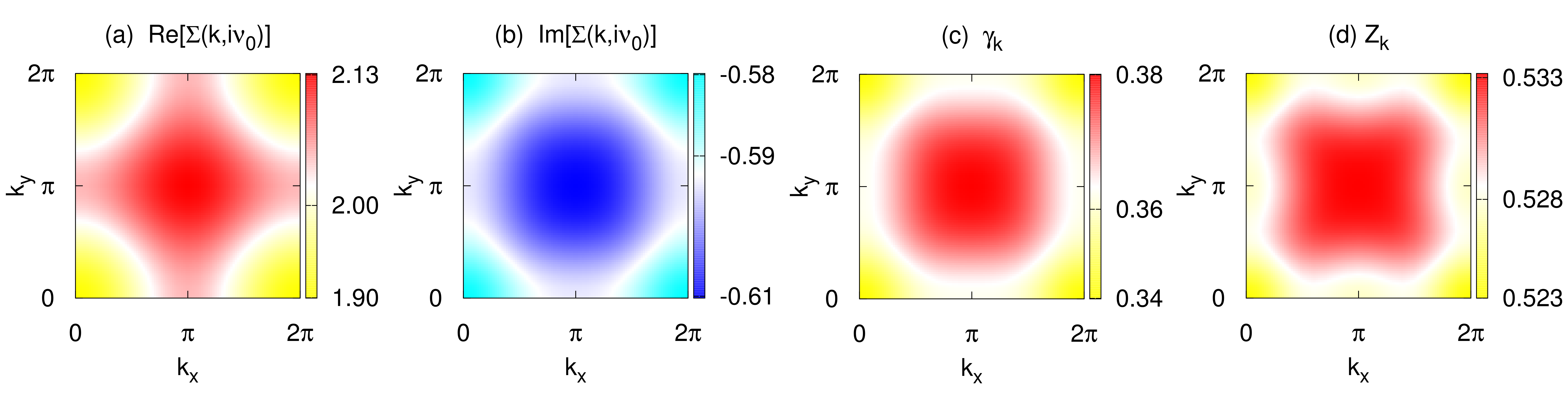}
    \end{flushleft}
    \caption{\label{FigDGASrVO3} Momentum-dependence in the
     $k_z$$=$$0$-plane of (a) the real and (b) imaginary part of the
     electronic self energy at the lowest Matsubara frequency
      ($i\nu_0$), computed in AbinitioD$\Gamma$A for the V-$t_\text{2g}$
      orbital $3d_{xy}$ of SrVO$_3$. The corresponding quasi-particle
      parameters (weight $Z_{\bf k}$) and scattering
        rate $\gamma_{\bf k}$, are reported in panels (c) and (d), respectively (reproduced from Ref.~\cite{Galler2016}).}
  \end{figure}

The application of the AbinitioD$\Gamma$A to SrVO$_3$ has allowed
one of the first non-perturbative analyses of the {\sl momentum}-dependence of the self energy,
 and of the spectral function in this
compound. 
The AbinitioD$\Gamma$A calculation foots on a DMFT solution of a realistic three-band Hubbard model for the t$_{2g}$-orbitals of vanadium.
The latter uses a static Hubbard $U=5.0$eV and Hund's $J=0.75$eV in the rotationally-invariant Kanamori parametrization.
The computation of two-particle quantities resorts to a recent worm algorithm \cite{Gunacker15}.
A sample of the AbinitioD$\Gamma$A results, for the $d_{xy}$-orbital, is
shown in Fig.~\ref{FigDGASrVO3}, where
 Re$\Sigma({\bf k}, i\nu)$ and  Im$\Sigma({\bf k}, i\nu)$ at the lowest
Matsubara-frequency ($i\nu=i\nu_0=\pi T$) are reported, as well as the related quasiparticle parameters ($Z_{\bf k}$) and
($\gamma_{\bf k}$) extracted from a low-frequency expansion of Im$\Sigma({\bf k}, i\nu_n)$.
These AbinitioD$\Gamma$A results show that that a sizable momentum
dependence does appear in the electronic self energy of SrVO$_3$
even if---as in this case---non-local {\it interactions} are neglected.
These effects thus correct
the purely local DMFT results. Interestingly, this {\bf k}-dependence is mostly confined to  Re$\Sigma({\bf k}, i\nu_0)$
(Fig.~\ref{FigDGASrVO3}a), where one observes a momentum differentiation
larger than $0.2$eV, which, however, does not directly mirror the
shape of the underlying Fermi-surface. At the same time, the  overall
{\bf k}-dependence of Im $\Sigma({\bf k}, i\nu_0)$, and the related
quasi-particle  coefficient $Z_{\bf k}$, $\gamma_{\bf k}$ (Fig.~\ref{FigDGASrVO3}b-d) is definitely much weaker (e.g., $Z_{\bf k}$ varies
less than $2\%$ over the whole Brillouin zone, with an average value
slightly increased w.r.t.\ DMFT).  We note that this behavior matches
rather well the conclusions of the space-time separation emerging from
previous single  band D$\Gamma$A  calculations \cite{Schaefer2015} discussed above.
Moreover, going beyond the single-orbital framework, it is also worth
emphasizing that a correlation between the momentum and orbital-dependence
is found: the strength of the {\bf k}-dependence, as computed in Ref.~\cite{Galler2016}, displays the same trends for the {\sl orbital}
dependence of $\Sigma$. The latter was found, indeed,  to be much more
pronounced for Re$\Sigma({\bf k}, i\nu_0)$,  than for the other
quantities shown in Fig.~\ref{FigDGASrVO3}.


\section{Conclusions}
\label{Sec:conclusion}

One of the  main challenges  in computational materials science is to predict the properties of materials for which the standard DFT-based methods are not applicable. Present DFT functionals are not reliable if the screening of the electron-electron interaction over different length- and time-scales is  insufficient to approximate the electronic exchange and correlation with the common LDA and GGA functionals. This is  the case for some correlated semiconductors, most transition metal oxides, as well as  heavy fermions.
In this paper, we have reviewed the forefront of methodological progress towards a full {\sl ab initio} treatment of  electronic exchange and correlations beyond DFT, and its state-of-the art merger with DMFT. These progresses range from the inclusion of the frequency dependencies in Hedin's {\it GW}-scheme for realistic calculations of large systems, to the implementation of a corresponding {\it GW}+DMFT algorithm that lifts the quasi-particle approximation in the scheme suggested by van Schilfgaarde and Kotani, and, eventually, to the treatment of  correlations {\sl beyond} the purely local description of DMFT by means of the AbinitioD$\Gamma$A. 
Such advances are of extreme importance, because the new algorithms are conceived to be able to treat {\sl all} classes of materials, independently of the strength and the range of the screened electronic interaction in the specific compound. While such ambitious goals will certainly require further work in the next decade, the examples we have selected in this work to illustrate the applications of the different method developments already show a promising trend.

Specifically, we have started by discussing the outline of the new $GW$ implementation in the Vienna {\it ab initio} simulation Package (VASP), written for massively parallel computations and working---for the first time within the VASP package---on the imaginary time/frequency axis. This allows not only the calculation of full dynamical information at the $GW$ level, but also opens the road  for the implementation of a self-consistency at the $GW$ level---beyond the quasi-particle approximation by  van Schilfgaarde and Kotani. The applicability of the new implementation has been demonstrated with a calculation of the testbed correlated metal SrVO$_3$. The progress in the $GW$ part are also pivotal for allowing a more natural and precise interfacing with DMFT-based algorithms. In particular, after reviewing the generic scheme of the {\it GW}+DMFT, where the non-local, but perturbative {\it GW}-exchange and correlations are supplemented with the purely local, but non-perturbative ones of DMFT, we have shown self energy results obtained
with the $G_0W_0$+DMFT merger of the VASP and the w2dynamics codes, again for the prototype material SrVO$_3$.
In particular, numerical results obtained by means of different levels of refinement (e.g., retaining/neglecting the dynamical structure of the screened interaction in the DMFT part) have been presented and critically analyzed.

Finally, we have reviewed the most general algorithmic framework in which even the limitations of {\it GW}+DMFT can be overcome, i.e., the AbinitioD$\Gamma$A approach. This  diagrammatic scheme starts with a local irreducible vertex  ($\Gamma_{\rm irr}$) obtained by using an impurity-solver (such as w2dynamics) and supplements it with the bare non-local Coulomb interaction. From this starting vertex,  ladder diagrams (or for a few orbitals parquet diagrams)
 are constructed, yielding non-local self energies and correlation functions.
 While one can easily obtain---within the AbinitioD$\Gamma$A formalism---all previously discussed approaches ({\it GW}, DMFT, and {\it GW}+DMFT) as limiting cases,
it also includes non-local correlations beyond these schemes, such as spin fluctuations.
 For a simple, one-band Hubbard model, we have recapitulated the unexpected properties found at the magnetic QCP in three dimensions, and discussed the space-time separability of the self energy.
For realistic multi-orbital materials calculations, we have shown the very first AbinitioD$\Gamma$A results for SrVO$_3$, and discussed the corrections found w.r.t. DMFT.
Indeed, we evidenced a sizable momentum-dependence in the SrVO$_3$ self-energy even for purely {\it local} interactions---an effect well beyond {\it GW} approaches \cite{Schaefer2015}.
Instead, the momentum dependence in our {\it GW}+DMFT results is almost exclusively propelled by exchange contributions originating from {\it non-local} interactions---thus far omitted in our AbinitioD$\Gamma$A calculations. As a consequence, the results of AbinitioD$\Gamma$A and {\it GW}+DMFT cannot yet be directly compared. Calculations that include the non-local interaction in AbinitioD$\Gamma$A 
are under way.

As it is typical in physics, and  especially true in the case of new algorithmic developments, a significant amount of future work will be inspired by the progress we have reviewed in this paper. In particular, the new method enhancements pave the way towards the implementation of  a fully self-consistent, frequency-dependent $GW$ scheme in VASP, while the frequency-dependent treatment of {\sl both} the $GW$ self energy and the local dynamic interaction of DMFT represents an important step towards the realization of a globally self-consistent {\it GW}+DMFT merger of the VASP and the w2dynamic codes. Eventually, the first successful applications of  AbinitioD$\Gamma$A for treating strong non-local correlations beyond {\it GW}+DMFT will encourage further efforts towards a new {\sl standard} of {\it ab initio} materials science calculations for correlated electron systems.

\bigskip

\section{Acknowledgments}
We thank   F.\ Aryasetiawan, S.\ Biermann, L.\ Boehnke, M.\ Casula, A.\ Galler, P.\ Gunacker, A.\ Hausoel, M.\ Kaltak, G.\ Li, A.\ Lichtenstein, T.\ Miyake, A.\ van Roekeghem, G.\ Rohringer, G.\ Sangiovanni, T.\ Sch{\"a}fer,   C.\ Taranto, P.\ Thunstr{\"o}m, and M.\ Wallerberger for fruitful discussions and cooperations, and in particular D.\ Springer who crucially contributed to the  presented {\it GW}+DMFT calculations using VASP and w2dynamics. Financial support is acknowledged  from the Austrian Science Fund (FWF) through the project  I 1395-N26 as part of the  research unit FOR 1346
of the Deutsche  Forschungsgemeinschaft (DFG).
P. Liu is grateful to the China Scholarship Council (CSC)-FWF Scholarship Program.



\end{document}